\documentclass[aps,prl,twocolumn,superscriptaddress]{revtex4-2}

\usepackage{multirow}
\usepackage{graphicx}
\usepackage{bm}
\usepackage{amsmath,amssymb}
\usepackage{hyperref}
\usepackage{lipsum}
\usepackage{color}

\usepackage{bm}
\usepackage{textcase}
\usepackage{textcomp}
\usepackage{braket}
\usepackage{setspace}
\usepackage{mathtools}
\usepackage[markup=default]{changes}

\definecolor{colorref}{rgb}{0.0, 0.408, 0.647}
\definecolor{grey}{rgb}{0.95, 0.95, 0.95}
\hypersetup{
	colorlinks = true,
	allcolors = colorref
}

\newcommand{\SIQSE}{\affiliation{1}{Shenzhen Institute for Quantum Science and Engineering, Southern University of Science and Technology, Shenzhen 518055, China}}
\newcommand{\PKU}{\affiliation{2}{School of Physics, Peking University, Beijing 100871, China}}
\newcommand{\IQA}{\affiliation{3}{International Quantum Academy, Shenzhen 518048, China}}
\newcommand{\GDKL}{\affiliation{4}{Guangdong Provincial Key Laboratory of Quantum Science and Engineering, Southern University of Science and Technology, Shenzhen 518055, China}}

\newcommand{\HFNL}{\affiliation{6}{
Shenzhen Branch, Hefei National Laboratory, Shenzhen 518048, China}}
\newcommand{\NXU}{\affiliation{7}{
School of Physics, Ningxia University, Yinchuan 750021, PR China}}
\newcommand{\ICT}{\affiliation{8}{
State Key Lab of Processors, Institute of Computing Technology, Chinese Academy of Sciences, Beijing 100190, China}}
\newcommand{\CETQC}{\affiliation{9}{
School of Computer Science and Technology, University of Chinese Academy of Sciences, Beijing 100049, China}}

\begin{document}

\title{Logical operations with a dynamical qubit in Floquet-Bacon-Shor code}

\date{\today}
\author{Xuandong Sun}
\thanks{These authors contributed equally to this work.}
\affiliation{\SIQSE}\affiliation{\IQA}\affiliation{\GDKL}

\author{Longcheng Li}
\thanks{These authors contributed equally to this work.}
\affiliation{\ICT}\affiliation{\CETQC}

\author{Zhiyi Wu}
\thanks{These authors contributed equally to this work.}
\affiliation{\PKU}\affiliation{\IQA}

\author{Zechen Guo}
\affiliation{\SIQSE}\affiliation{\IQA}\affiliation{\GDKL}

\author{Peisheng Huang}
\affiliation{\NXU}\affiliation{\IQA}

\author{Wenhui Huang}
\affiliation{\SIQSE}\affiliation{\IQA}\affiliation{\GDKL}

\author{Qixian Li}
\affiliation{\SIQSE}\affiliation{\IQA}\affiliation{\GDKL}

\author{Yongqi Liang}
\affiliation{\SIQSE}\affiliation{\IQA}\affiliation{\GDKL}

\author{Yiting Liu}
\affiliation{\SIQSE}\affiliation{\IQA}\affiliation{\GDKL}

\author{Daxiong Sun}
\affiliation{\SIQSE}\affiliation{\IQA}\affiliation{\GDKL}

\author{Zilin Wang}
\affiliation{\NXU}\affiliation{\IQA}

\author{Changrong Xie}
\affiliation{\SIQSE}\affiliation{\IQA}\affiliation{\GDKL}

\author{Yuzhe Xiong}
\affiliation{\SIQSE}\affiliation{\IQA}\affiliation{\GDKL}

\author{Xiaohan Yang}
\affiliation{\SIQSE}\affiliation{\IQA}\affiliation{\GDKL}

\author{Jiajian Zhang}
\affiliation{\SIQSE}\affiliation{\IQA}\affiliation{\GDKL}

\author{Jiawei Zhang}
\affiliation{\SIQSE}\affiliation{\IQA}\affiliation{\GDKL}

\author{Libo Zhang}
\affiliation{\SIQSE}\affiliation{\IQA}\affiliation{\GDKL}

\author{Zihao Zhang}
\affiliation{\SIQSE}\affiliation{\IQA}\affiliation{\GDKL}

\author{Weijie Guo}
\affiliation{\IQA}

\author{Ji Jiang}
\affiliation{\SIQSE}\affiliation{\IQA}\affiliation{\GDKL}

\author{Song Liu}
\affiliation{\SIQSE}\affiliation{\IQA}\affiliation{\GDKL}\affiliation{\HFNL}

\author{Xiayu Linpeng}
\affiliation{\IQA}

\author{Jingjing Niu}
\affiliation{\IQA}\affiliation{\HFNL}

\author{Jiawei Qiu}
\affiliation{\IQA}

\author{Wenhui Ren}
\affiliation{\IQA}

\author{Ziyu Tao}
\affiliation{\IQA}

\author{Yuefeng Yuan}
\affiliation{\IQA}

\author{Yuxuan Zhou}
\affiliation{\IQA}

\author{Ji Chu}
\email{jichu@iqasz.cn}
\affiliation{\IQA}

\author{Youpeng Zhong}
\email{zhongyp@sustech.edu.cn}
\affiliation{\SIQSE}\affiliation{\IQA}\affiliation{\HFNL}

\author{Xiaoming Sun}
\email{sunxiaoming@ict.ac.cn}
\affiliation{\ICT}\affiliation{\CETQC}

\author{Dapeng Yu}
\email{yudapeng@iqasz.cn}
\affiliation{\IQA}\affiliation{\HFNL}

\begin{abstract}
Quantum error correction (QEC) protects quantum systems against inevitable noises and control inaccuracies, providing a pathway towards fault-tolerant (FT) quantum computation. Stabilizer codes, including surface code and color code, have long been the focus of research and have seen significant experimental progress in recent years.
Recently proposed time-dynamical QEC, including Floquet codes and generalized time-dynamical code implementations, opens up new opportunities for FT quantum computation. By employing a periodic schedule of low-weight parity checks, Floquet codes can generate additional dynamical logical qubits, offering enhanced error correction capabilities and potentially higher code performance.
Here, we experimentally implement the Floquet-Bacon-Shor code on a superconducting quantum processor. We encode a dynamical logical qubit within a $3\times 3$ lattice of data qubits, alongside a conventional static logical qubit.
We demonstrate FT encoding and measurement of the two-qubit logical states, and stabilize these states using repeated error detection. We showcase universal single-qubit logical gates on the dynamical qubit. Furthermore, by implementing a logical CNOT gate, we entangle the dynamical and static logical qubits, generating an error-detected logical Bell state with a fidelity of 75.9\%. Our results highlight the potential of Floquet codes for resource-efficient FT quantum computation.
\end{abstract}

\maketitle

\begin{figure*}[t]
    \centering
    \includegraphics[width=\textwidth]{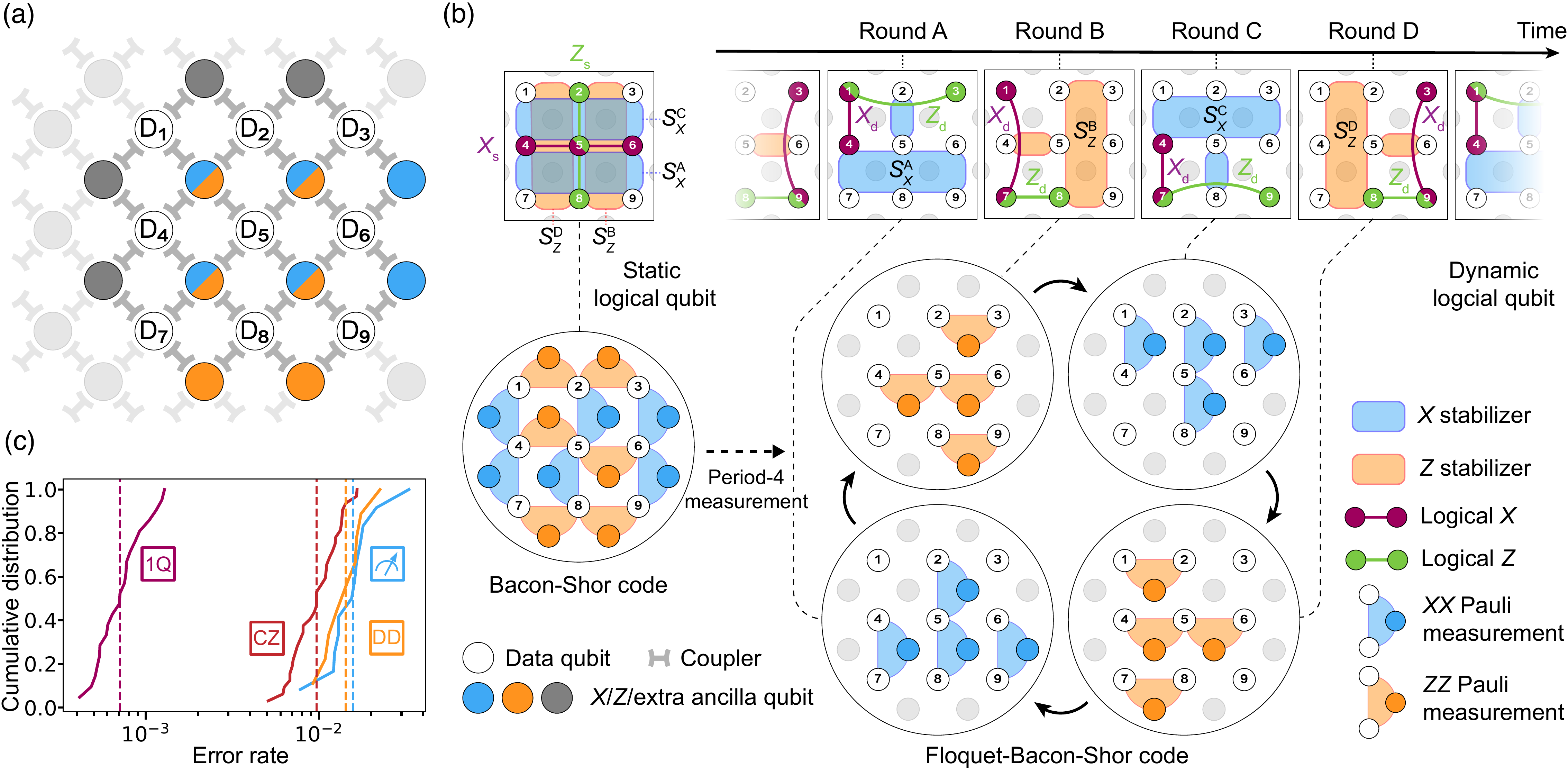}
    \caption{
    Implementation of the FBS code.
    (a) Schematic of the quantum device for the FBS code, consisting of 9 data qubits (white circles labeled as $D_i$, where i = 1 to 9) and 12 ancilla qubits (orange circles for $Z$ ancillas and blue circles for $X$ ancillas). The four central qubits are alternately used as $X$ and $Z$ ancillas in different stabilizer rounds. Extra ancilla qubits (grey circles) are used for circuit compression and the construction of logical CNOT gate. Each pair of qubits is connected by a tunable coupler (H shape).
    (b) Illustration of stabilizer measurements for the BS code (left) and the FBS code (right). The FBS code features period-4 stabilizer measurements, during which the logical operators $X_{\rm d}$ (purple lines) and $Z_{\rm d}$ (green lines) for the dynamical qubit evolve across different rounds. In contrast, the logical operators $X_{\rm s}$ and $Z_{\rm s}$ for the static qubit remain fixed and are identical in both codes. Weight-2 Pauli measurements are used to check the parity of the corresponding data qubits in the $X$ (blue) and $Z$ (orange) bases.
    (c) Cumulative distributions of simultaneous errors for single-qubit gates, CZ gates, ancilla qubit readout, and data qubit idle (with dynamical decoupling) during ancilla readout are depicted. The median error rates are 0.07\%, 0.97\%, 1.58\%, and 1.43\%, respectively, as indicated by the vertical lines.} \label{fig:main}
\end{figure*}

Quantum computers hold potential for solving problems intractable for classical systems~\cite{lloyd1996universal,shor1999polynomial,biamonte2017quantum}. 
However, their practical realization faces the fundamental challenge of quantum information fragility.
To address this challenge, quantum error correction (QEC) has been established as a critical framework~\cite{knill1997theory,nielsen2010quantum,terhal2015quantum}. By redundantly encoding logical qubits into larger systems with many physical qubits, QEC detects and corrects errors through repeated stabilizer measurements. 
QEC codes, including surface code~\cite{kitaev2003fault}, color code~\cite{bombin2006topological}, Bacon-Shor (BS) code~\cite{bacon2006operator}, and quantum LDPC codes~\cite{bravyi2024high}, have made significant experimental progress across various systems in recent years~\cite{egan2021fault,postler2022demonstration,ryan2024high,pogorelov2025experimental,paetznick2024demonstration,zhao2022realization,krinner2022realizing,google2023suppressing,bluvstein2024logical,abobeih2022fault}. 
Particularly in logical state preservation, exponential error suppression has been achieved by scaling a surface-code logical qubit in superconducting circuits~\cite{google}. Beyond state preservation, logical operations—such as arbitrary single-qubit gates and entangling gates between logical qubits—are essential for fault-tolerant (FT) quantum computing~\cite{fowler2012surface}. However, implementing these operations is generally challenging, especially in physical systems with local connectivity~\cite{marques2022logical,lacroix2024scaling,besedin2025realizing}.

Recently proposed time-dynamical QEC, including Floquet codes~\cite{hastings2021dynamically,haah2022boundaries,vuillot2021planar,gidney2021fault,gidney2022benchmarking,zhang2022x,aasen2022adiabatic,wootton2022measurements,davydova2023floquet,townsend2023floquetifying,sullivan2023floquet,ellison2023floquet,fahimniya2023fault,paetznick2023performance,aasen2023measurement,alam2024dynamical,bauer2024topological,higgott2024constructions,hilaire2024enhanced,fu2024error,setiawan2024tailoring,kesselring2024anyon,dua2024engineering,hesner2025using} and generalized time-dynamical code implementations~\cite{gottesman2022opportunities,mcewen2023relaxing,gidney2023new,delfosse2023spacetime,shaw2024lowering,eickbusch2024demonstrating}, opens up new opportunities for FT quantum computation by transcending the usual paradigm of stabilizer codes or subsystem codes, where logical information is encoded in a fixed subspace.
The Floquet codes hold the advantage of requiring only low-weight Pauli
measurements, typically of weight-2~\cite{hastings2021dynamically}, which are naturally compatible with physical architectures supporting native two-qubit measurements~\cite{riste2013deterministic,aasen2025roadmap}.
Moreover, by periodically encoding logical information over time, Floquet codes enable additional dynamical logical qubits compared to their parent codes, offering greater flexibility and potentially higher code performance~\cite{paetznick2023performance}.

In this work, we implement the Floquet-Bacon-Shor (FBS) code~\cite{alam2024dynamical} on a superconducting quantum processor. As a minimal, distance-2 implementation of a Floquet code, it requires only square-lattice connectivity and is thus readily implementable on contemporary hardware.
Compared to the conventional BS code, the FBS code additionally encodes one dynamical logical qubit, enabling two-qubit logical operations.
We demonstrate FT encoding and measurement of the two-qubit logical states, and stabilize these states using multi-round stabilizer measurements. We showcase FT Pauli gates and non-fault-tolerant (nFT) arbitrary rotation gates on the dynamical qubit. Furthermore, we achieve a logical Bell state with a fidelity of $75.9\%$ by entangling the dynamical and static logical qubits via a logical CNOT gate, with the gate fidelity assessed at $84.1\%$ through logical quantum process tomography (LQPT). Our results highlight the potential of Floquet codes for scalable and resource-efficient QEC.

\begin{figure*}[t]
    \centering
    \includegraphics[width=1.0\textwidth]{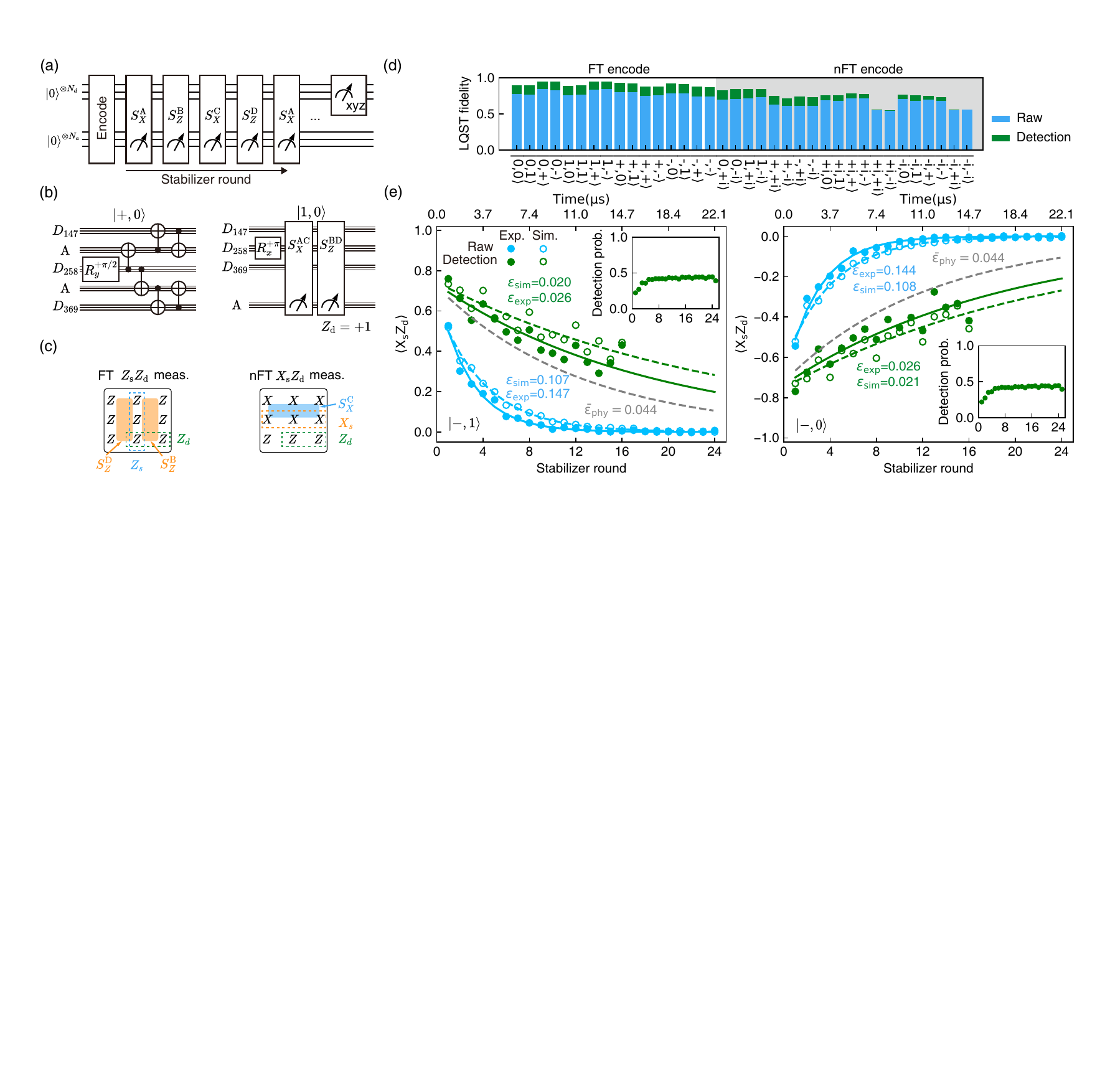}
    \caption{
    Encoding, stabilization, and measurement of two-qubit logical states in the FBS code.
    (a) Simplified circuit for encoding, stabilizing and measuring two-qubit logical states. All physical qubits are initialized to the ground state before the circuit. Period-4 stabilizer measurements are alternately performed following the encoding circuit.
    (b) Encoding circuits for logical Pauli states $|+,0\rangle$ and $|1,0\rangle$. BS stabilizer measurements are included in the latter case, with the dynamical qubit state post-selected to be $|0\rangle$. The notation $D_{147}$ denotes the three data qubits $D_1$, $D_4$, and $D_7$, and so on.
    The CNOT symbol denotes transversal CNOT gates between data and ancilla qubit groups, subject to connectivity constraints.
    (c) Example logical measurement circuits after stabilizer $S_Z^{\rm D}$: FT $ZZ$ measurement (left) and nFT $XZ$ measurement (right). Logical operators are indicated within the dashed-line boxes.
    Measurement outcomes of data qubits are used to calculate the values of the $X$ (blue) and $Z$ (orange) stabilizers, enabling error detection.
    (d) LQST fidelities of all 36 two-qubit Pauli states.
    The 16 states $\{|0\rangle,|1\rangle,|+\rangle,|-\rangle\}^{\otimes 2}$ are fault-tolerantly prepared and exhibit higher fidelity, as shown on the left side (white background).
    The blue portions indicate the raw fidelities (i.e., without using syndrome data), while the green portions highlight the fidelity improvement by error detection (i.e., with syndrome data post-selected).
    (e) Experimentally measured (filled symbols) and simulated (open symbols) expectation values of $X_{\rm s}Z_{\rm d}$ operator versus stabilizer round, for the encoded states $|-,1\rangle$ (left) and $|-,0\rangle$ (right). The solid and dashed lines represent exponential fits. Each stabilizer round lasts 920~ns, comprising 720~ns for ancilla readout and 200~ns for single-qubit and two-qubit gates. The cyan and green colors correspond to the raw data and error-detected data, respectively. The dashed grey lines indicate the average error rates of physical two-qubit states. Due to the exponential decrease in the retained data during error detection, the analysis is limited to 16 rounds, with a total of 100,000 experimental shots. The insets show the error detection probabilities for the weight-6 stabilizers over 24 measurement rounds, with the final round calculated from the logical measurements.
    } \label{fig2_state_preservation}
\end{figure*}

The FBS code we demonstrate is a generalization of the $[[9,1,3]]$ BS code~\cite{bacon2006operator}, which encodes one logical qubit in a $3 \times 3$ square lattice, as illustrated in Fig.~\ref{fig:main}(a). 
Each labeled vertex represents a data qubit, and other colored vertices represent ancilla qubits for weight-2 Pauli measurements.
The static logical qubit is defined within a fixed subsystem by the operators: \begin{equation}
    X_{\rm s} = X_4 X_5 X_6, \quad Z_{\rm s} = Z_2 Z_5 Z_8, \quad Y_{\rm s} = i X_{\rm s} Z_{\rm s}.
    \label{eq:static_qubit_operator}
\end{equation} 
The BS code is stabilized by weight-2 Pauli measurements~\cite{bacon2006operator}, whose products yield four weight-6 stabilizers $S_X^{\text{A}}$, $S_Z^{\text{B}}$, $S_X^{\text{C}}$, and $S_Z^{\text{D}}$, enabling the detection and correction of any single-qubit error. 
Since these Pauli measurements do not always commute, the Hilbert space of the 9 physical qubits consists of 4 degrees of freedom fixed by the stabilizers, 1 degree for logical information, and 4 additional degrees corresponding to the gauge space, which does not encode logical information. 
By imposing a period-4 ordering on the weight-2 Pauli measurements, the gauge space dynamically encodes an additional logical qubit, thereby forming the FBS code. 
The logical Pauli operators $X_{\text{d}}$ and $Z_{\text{d}}$ of the dynamically encoded qubit evolve periodically during the stabilizer measurements, as indicated by the purple and green lines in Fig.~\ref{fig:main}(b). 
Throughout this paper, the initial operator definitions are consistent with stabilizer round A. 
Although the $3 \times 3$ system size is insufficient for error correction of the dynamical qubit, the stabilizers extracted from the measurements are sufficient to detect any single-qubit error. 
Within the framework of error detection, an operation is considered FT if any single-qubit fault produces a non-trivial syndrome and is thereby detectable~\cite{tomita2014low}.

We implement the FBS code on a 21-qubit subsystem of a 66-transmon superconducting processor~\cite{koch2007charge} with 110 tunable couplers~\cite{yan2018tunable}.
The $N_d=9$ data qubits, labeled as $D_i$ ($i=1$ to 9), form a $3 \times 3$ array and are interlaced with $X$ and $Z$ ancilla qubits for stabilizer measurements. The cumulative distributions of simultaneous gate errors and readout errors~\cite{braginsky1996quantum} are illustrated in Fig.~\ref{fig:main}(c). The experimental setup, device parameters and calibration methods are detailed in the Supplementary Material~\cite{SM2025}.

\begin{figure}[t]
    \centering
    \includegraphics[width=\columnwidth]{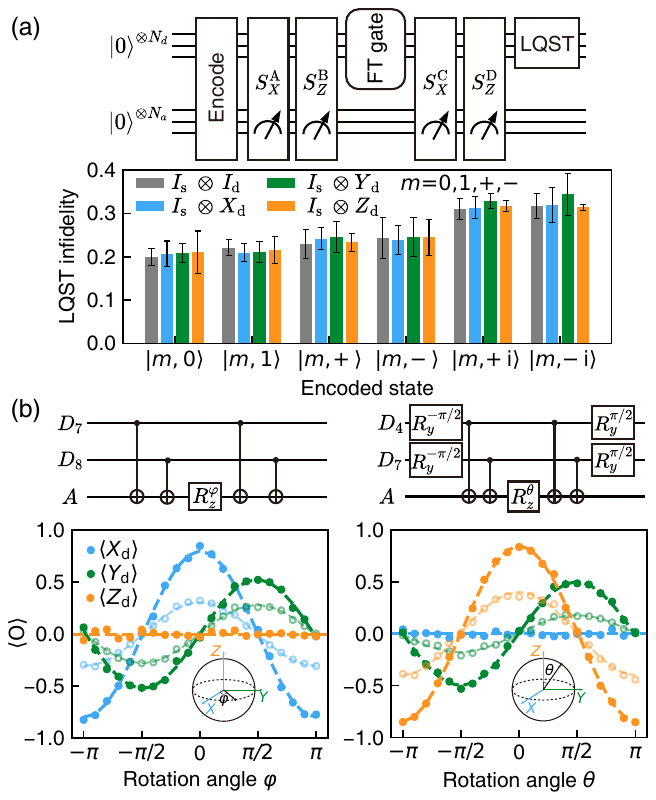}
    \caption{Logical gates on the dynamical qubit.
    (a) FT Pauli gates on the dynamical qubit. Transversal Pauli gates are applied to the data qubits and inserted between stabilizers $S_Z^{\rm B}$ and $S_X^{\rm C}$, as shown at the top.
    The final state is characterized using LQST. The state infidelities with error detection after applying the Pauli gates on different encoded states are shown below the circuit. For each dynamical qubit state, the infidelity is averaged over 4 static qubit states \{$|0\rangle$, $|1\rangle$, $|+\rangle$, $|-\rangle$\}. Error bars correspond to 95\% confidence intervals.
    (b) Logical rotation gates on the dynamical qubit around the $Z_{\rm d}$-axis (left) and $X_{\rm d}$-axis (right). The measurement expectation values of the operators $X_{\rm d}$ (cyan), $Y_{\rm d}$ (green) and $Z_{\rm d}$ (orange) are shown as functions of rotation angles $\varphi$ and $\theta$. 
    The gate circuits (shown at the top) for rotations around the $Z_{\rm d}$-axis ($X_{\rm d}$-axis) are inserted between stabilizers $S_Z^{\rm B}$ ($S_X^{\rm C}$) and $S_X^{\rm C}$ ($S_Z^{\rm D}$). The filled (open) symbols represent data with (without) error detection in the four stabilizer rounds and the final logical measurement. The dashed curves correspond to trigonometric fits.
    } \label{fig3_ft_gate}
\end{figure}

The simplified circuit for logical state encoding, period-4 stabilization, and measurement is shown in Fig.~\ref{fig2_state_preservation}(a), with all physical qubits initialized to ground state~\cite{johnson2012heralded,yang2024coupler}. 
We prepared 36 two-qubit Pauli eigenstates $\{|0\rangle, |1\rangle, |+\rangle, |-\rangle, |{+}i\rangle, |{-}i\rangle\}^{\otimes 2}$, adopting the notation $|$static qubit, dynamical qubit$\rangle$.
$ZX$/$XZ$-basis states (e.g., $|+,0\rangle$) are prepared fault-tolerantly using BS code methods~\cite{egan2021fault}, while $XX$/$ZZ$-basis states (e.g., $|1,0\rangle$) are prepared via stabilizer measurements and post-selection (Fig.~\ref{fig2_state_preservation}(b)). States involving $\ket{\pm i}$ are prepared non-fault-tolerantly~\cite{SM2025}.
Logical measurements are performed by measuring the data qubits in their appropriate bases, as illustrated by the example circuits in Fig.~\ref{fig2_state_preservation}(c), and the logical outcome is determined by parity calculation.
The $Z_{\rm s}Z_{\rm d}$ measurement is FT because the measured outcomes allow for the calculation of stabilizers $S_Z^{\rm D}$ and $S_Z^{\rm B}$, enabling the detection of any single-qubit error. The $X_{\rm s}Z_{\rm d}$ measurement is nFT but $S_X^{\rm C}$ value is calculated to detect phase-flip errors and measurement errors. The fidelities of 36 encoded states are assessed via logical quantum state tomography (LQST), where 9 logical bases $\{Z,X,Y\}^{\otimes 2}$ are measured~\cite{SM2025}. Error detection during the logical measurements enhances the fidelities, as shown by the green portions in Fig.~\ref{fig2_state_preservation}(d).

We perform period-4 stabilizer measurements to preserve the encoded two-qubit logical states. In each stabilizer round, 4 weight-2 Pauli measurements are conducted. Each measurement result is compared to the result of the same stabilizer type from four rounds earlier, with the first four rounds compared to the initial encoding values. The preservation results for the encoded states $|-,0\rangle$ and $|-,1\rangle$ are presented in Fig.~\ref{fig2_state_preservation}(e), showing the measured expectation values of the operator $X_{\rm s}Z_{\rm d}$. With error detection, the logical error rates per round ($\varepsilon_{\rm exp}$), extracted by fitting the exponential decay of the expectation values, are reduced from 14.7\% and 14.3\% to 2.6\% and 2.5\%, respectively. The error rates are compared to those of physical two-qubit states (dashed grey lines), with $\overline{\varepsilon}_{\rm phy} = 1 - \exp(-\tau_{\rm exp}/\overline{T_{2}^{\rm echo}})/2 - \exp(-\tau_{\rm exp}/\overline{T_{1}})/2=4.4\%$.
Here, $\tau_{\rm exp} =920$~ns is the duration of a stabilizer round, $\overline{T_{2}^{\rm echo}}=11.7~\mu s$ is the average spin-echo dephasing time, and $\overline{T_{1}} = 77.1~\mu s$ is the average energy relaxation time of the physical qubits.
Numerical simulation results based on monitored gate and readout errors are also shown for comparison~\cite{gidney2021stim}. The larger error rates observed in experiments may arise from leakage-induced correlated errors~\cite{SM2025,fowler2014quantifying,mcewen2021removing,lacroix2023fast,bultink2020protecting,miao2023overcoming,marques2023all,yang2024coupler}.   
In both the experimental and simulation data, round-dependent oscillations are observed, with higher expectation values occur every four rounds, corresponding to the completion of a full error detection cycle.

\begin{figure}[t]
    \centering
    \includegraphics[width=\columnwidth]{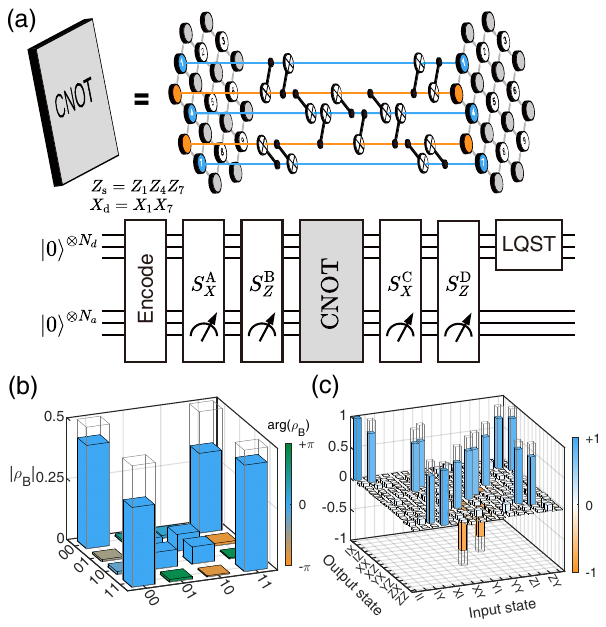}
    \caption{Logical CNOT gate between the dynamical and static qubits.
    (a) The CNOT circuit (top) is inserted between stabilizers $S_Z^{\rm B}$ and $S_X^{\rm C}$ in the overall circuit (bottom).
    For Bell state generation, the logical qubits are initialized to $|+,0\rangle$ during the encoding process. For LQPT, 16 linearly independent logical cardinal states: $\{|0\rangle, |1\rangle, |-\rangle, |{-}i\rangle \}^{\otimes 2}$ are encoded.
    (b) Error-detected logical density matrix of the generated logical Bell state, obtained with four stabilizer rounds. The logical state fidelity is 75.9\% with error detection and 38.8\% without error detection.
    The wireframes indicate the ideal values.
    (c) Extracted LPTM $R_{\rm exp}$ of the CNOT gate, with a process fidelity $F_{\rm p} = 80.2\%$.  
    } \label{fig4_cnot}
\end{figure}

We demonstrate universal logical gates on the dynamical qubit, including FT Pauli gates and nFT arbitrary rotation gates. For logical gates on the static qubit, refer to the Supplementary Material~\cite{SM2025} and reference~\cite{egan2021fault}.
Pauli gates are implemented transversally via single-qubit gates on corresponding data qubits, inserted between stabilizers $S_Z^{\rm B}$ and $S_X^{\rm C}$ (Fig.~\ref{fig3_ft_gate}(a)).
LQST characterization reveals that the average state infidelities after applying the gates $I_{\rm s} \otimes X_{\rm d}$, $I_{\rm s} \otimes Y_{\rm d}$, and $I_{\rm s} \otimes Z_{\rm d}$ are $0.255\pm0.011$, $0.263\pm0.013$, and $0.256\pm0.011$, respectively. These values closely match the state infidelity without gate inserted ($0.253\pm0.011$), confirming that the FT Pauli gates induce negligible errors.

Arbitrary logical rotations are constructed using ancilla qubits. The $R_Z^\varphi$ gate is realized by inserting a physical $R_z^{\varphi}$ gate on an ancilla between four CNOTs on $D_7$ and $D_8$, placed between stabilizers $S_Z^{\rm B}$ and $S_X^{\rm C}$, as illustrated at the top of Fig.~\ref{fig3_ft_gate}(b). $R_X^\theta$ is constructed in a similar way and is inserted between $S_X^{\rm C}$ and $S_Z^{\rm D}$. 
Sweeping angles $\varphi$ and $\theta$ and measuring $X_{\rm d}$, $Y_{\rm d}$, and $Z_{\rm d}$ yields the results in Fig.~\ref{fig3_ft_gate}(b), where filled (open) symbols denote data with (without) error detection. The reduced $Y_{\rm d}$ amplitude under error detection results from its nFT measurement. 
Without error detection, $R_X^\theta$ gate exhibits a further suppressed $Y_{\rm d}$ amplitude compared to $Z_{\rm d}$, a consequence of enhanced phase-flip error propagation through this nFT gate. Data post-processing is used to avoid real-time feedback~\cite{SM2025}.

Finally, we demonstrate entanglement between the dynamical and static qubits using a logical CNOT gate. The gate circuit, inserted between stabilizers $S_Z^{\rm B}$ and $S_X^{\rm C}$, involves 14 entangling gates acting on 3 data qubits and 2 ancilla qubits, as illustrated in Fig.~\ref{fig4_cnot}(a). 
The circuit is derived by using the static qubit operator $Z_{\rm s}=Z_1Z_4Z_7$ (equivalent to $Z_{\rm s}$ in Eq.~\ref{eq:static_qubit_operator}) to control the dynamical qubit operator $X_{\rm d}=X_1X_7$ after stabilizer $S_Z^{\rm B}$~\cite{SM2025}. 
Although the logical CNOT gate is nFT due to error propagation among the 3 data qubits, the circuit is designed to minimize such effects.

Applying the logical CNOT gate to $|+,0\rangle$ yields a logical Bell state $\frac{1}{\sqrt{2}}(|0,0\rangle + |1,1\rangle )$. The error-detected logical density matrix $\rho_{\rm B}$ is depicted in Fig.~\ref{fig4_cnot}(b), with a fidelity of 75.9\%. 
We benchmark the gate fidelity using LQPT with 16 input states $\{|0\rangle, |1\rangle, |-\rangle, |{-}i\rangle \}^{\otimes 2}$. 
For each input, two-qubit LQST is performed at the end of the circuit.
To account for state preparation and measurement (SPAM) errors, we characterize the input states separately using the same circuit without the CNOT gate~\cite{marques2022logical}. 
The resulting logical Pauli transfer matrix (LPTM) $R_{\rm exp}$ (Fig.~\ref{fig4_cnot}(c)) is constructed from the 16 input-output pairs~\cite{SM2025}. 
Comparing it with the ideal matrix $R_{\rm ideal}$ gives a process fidelity of $\mathcal{F}_{p} = {\rm Tr}(R_{\rm exp} \cdot R_{\rm ideal}) = 80.2\%$, which corresponds to a gate fidelity $\mathcal{F}_{g} = (\mathcal{F}_{p}/E+1)/(E+1)=84.1\%$ for the $E=4$-dimensional Hilbert space~\cite{chow2012universal}.
The LQPT fidelity suffers from the relatively low fidelity of encoded $|{-}i\rangle$ states. As an alternative estimate, we exclude these states and compute the ratio of the average output state fidelity to the average input state fidelity, which yields a gate fidelity of 93.6\%.

In conclusion, we implement the FBS code on a superconducting quantum processor, encoding both a dynamical logical qubit and a conventional static logical qubit. We demonstrate universal logical operations on the dynamical qubit, integrated with repeated quantum error detection, and entangle the dynamical and static logical qubits by constructing a logical CNOT gate. 
Future work may scale Floquet codes to larger lattices for dynamical qubit error correction.
However, the standard FBS code cannot provide a practical scaling due to lack of a threshold~\cite{alam2024dynamical}. Techniques such as code concatenation~\cite{aliferis2007subsystem, cross2007comparative} and measurement skipping~\cite{gidney2023less,alam2025bacon} could be adapted to enable effective error suppression at larger code distances.
Additionally, methods like lattice surgery~\cite{haah2022boundaries} and braiding twist defects~\cite{ellison2023floquet} may be integrated into Floquet codes to enable FT entanglement and non-Clifford gates with the dynamical logical qubit.

We thank Bujiao Wu for valuable discussion.
This work was supported by the Science, Technology and Innovation Commission of Shenzhen Municipality (KQTD20210811090049034), the Innovation Program for Quantum Science and Technology (2021ZD0301703), the National Natural Science Foundation of China (62325210, 12174178, 12374474, 123b2071).

The data that support the findings of this article are openly available~\cite{sun_2025_15783707}.

\bibliography{sn-bibliography}  
\bibliographystyle{apsrev4-2}

\end{document}


\title{Supplemental Material for ``Logical operations with a dynamical qubit in Floquet-Bacon-Shor code''}

\maketitle

\setcounter{equation}{0}
\setcounter{figure}{0}
\setcounter{table}{0}
\setcounter{page}{1}

\tableofcontents
\newpage

\section{Logical operator definition in the FBS code}

The FBS code is derived from its parent subsystem code, the $[[9,1,3]]$ BS code, which employs a $3\times 3$ square lattice of data qubits to encode one logical qubit. The gauge group $G$ of the BS code is generated by the following $12$ weight-$2$ gauge operators (refer to Fig.~1(a) in the main text for qubit
numbering):
\begin{align*}
    \langle &Z_1Z_2, Z_2Z_3, Z_4Z_5, Z_5Z_6, Z_7Z_8, Z_8Z_9, \\
&X_1X_4, X_2X_5, X_3X_6, X_4X_7, X_5X_8, X_6X_9 \rangle .
\end{align*}
By identifying the operators in $G$ that commute with all elements in $G$, we obtain $4$ weight-$6$ stabilizers:
\begin{align*}
    S_X^{\rm A}&=X_4X_5X_6X_7X_8X_9,\quad S_Z^{\rm B}=Z_2Z_3Z_5Z_6Z_8Z_9,\\
    S_X^{\rm C}&=X_1X_2X_3X_4X_5X_6,\quad S_Z^{\rm D}=Z_1Z_2Z_4Z_5Z_7Z_8.
\end{align*}
The values of the above stabilizers can be obtained by measuring all $12$ gauge operators and multiplying the corresponding measurement outcomes. By identifying the operators that commute with all elements in $G$ but are not themselves in $G$, we obtain the logical operators of the static logical qubit, as defined in Eq.~1 in the main text.

The FBS code is constructed by measuring the gauge operators in a period-$4$
schedule (see Fig.~1(b) in the main text), thereby encoding an
additional dynamical logical qubit. Up to a $\pm 1$ sign, the logical operator of the dynamical qubit evolves according to the period-$4$ schedule, as shown in Table~\ref{tab:dynamical}. In the remainder of this section, we omit the superscript ``Q'' if it is clear from the context.
\begin{table}[htbp]
    \begin{tabular}{|c||c|c|c|c|}
        \hline
        Round Q & A & B & C & D \\
        \hline
        $X_{\rm d}^{\rm Q}$ & $X_1X_4$ & $X_1X_7$ & $X_4X_7$ & $X_3X_9$\\
        \hline        
        $Z_{\rm d}^{\rm Q}$ & $Z_1Z_3$ & $Z_7Z_8$ & $Z_7Z_9$ & $Z_8Z_9$\\         
        \hline        
        $Y_{\rm d}^{\rm Q}$ & $Y_1Z_3X_4$ & $X_1Y_7Z_8$ & $X_4Y_7Z_9$ & $X_3Z_8Y_9$\\         
        \hline
    \end{tabular}
    \caption{Expressions for \emph{unsigned} dynamical logical operators.}
    \label{tab:dynamical}
\end{table}
For Pauli operator $P\in\{X,Y,Z\}$, its sign at round $r$ is given by formula
$\Gamma_{P}^{(r)}=\prod_{i=1}^{r}\gamma_P^{(i)}$, where $\gamma_P^{(i)}$ is
defined in Table~\ref{tab:phase}. Combining all of the above, the logical $P$ operator of the dynamical qubit at round $r$ is given by
\(
P_{\rm d}^{(r)}=\Gamma_{P}^{(r)} P_{\rm d}^{\rm Q},
\)
where \( {\rm Q} = {\rm A}, {\rm B}, {\rm C}, {\rm D} \) for \( r \equiv 1, 2, 3, 0 \pmod{4} \), respectively.

\begin{table}[htbp]
    \begin{tabular}{|c||c|c|c|c|}
        \hline
        $i~\mathrm{mod}~4$ & 1 & 2 & 3 & 0 \\
        \hline
        $\gamma_X^{(i)}$ & $x_{69}^{(i)}x_{25}^{(i)}S_X^{\rm C}$ & $x_{47}^{(i-1)}$ & $x_{14}^{(i)}$ & $x_{36}^{(i-1)}x_{58}^{(i-1)}S_{X}^{\rm A}$\\
        \hline
        $\gamma_Z^{(i)}$ & $z_{12}^{(i-1)}z_{56}^{(i-1)}S_Z^{\rm B}$ & $z_{23}^{(i)}z_{45}^{(i)}S_Z^{\rm D}$ & $z_{89}^{(i-1)}$ & $z_{78}^{(i)}$ \\ 
        \hline
        $\gamma_Y^{(i)}$ & \multicolumn{4}{|c|}{$\gamma_X^{(i)}\gamma_Z^{(i)}$} \\
        \hline
    \end{tabular}
    \caption{Definition of $\gamma^{(i)}_X, \gamma^{(i)}_Z, \gamma^{(i)}_Y$. At round
    $i$, $x_{uv}^{(i)}, z_{uv}^{(i)} \in \{\pm 1\}$ denotes the ancilla
    measurement outcome of weight-$2$ Pauli operators $X_uX_v$ and $Z_uZ_v$. $S_X^{\rm A},
    S_Z^{\rm B}, S_X^{\rm C}, S_Z^{\rm D} \in \{\pm 1\}$ refer to the initial values of the
    stabilizers determined by the encoding process. } \label{tab:phase}
\end{table}

\subsection{Logical operator evaluation using data-qubit measurement outcomes}

For the static qubit, the logical operator is calculated from the measurement outcomes of data qubits during logical measurement. The values of the static logical operators are evaluated by multiplying the measurement outcomes of the corresponding data qubits, i.e., 
\(
X_{\rm s} = x_4 x_5 x_6, \quad Z_{\rm s} = z_2 z_5 z_8, \quad Y_{\rm s} = x_4 x_6 y_5 z_2 z_8,
\)  
where \( x_i, y_i, z_i \in \{\pm 1\} \) denote the $X$, $Y$, and $Z$ measurement outcomes of data qubit \( D_i \), respectively.

Assume the logical measurement is performed after $r$ rounds of stabilizer measurement. For Pauli operator $P\in\{X,Y,Z\}$, the value of the dynamical logical operator $P^{\rm (r)}_{\rm d}= \Gamma_{P}^{(r)}P_{\rm d}$ is the product of two parts: the value of its unsigned operator $P_{\rm d}$ and the value of its sign $\Gamma_{P}^{(r)}$ at round $r$. Similar to the static logical operators, the value of $P_{\rm d}$ is obtained by multiplying the measurement outcomes of the corresponding data qubits (see Table~\ref{tab:dynamical}). The value of $\Gamma_{P}^{(r)}$ is calculated as
$\Gamma_{P}^{(r)}=\prod_{i=1}^{r}\gamma_{P}^{(i)}$, where $\gamma_{P}^{(i)}$ is determined by the ancilla measurement outcomes during encoding and stabilizer measurements (see Table~\ref{tab:phase}). Due to the presence of $\Gamma_{P}^{(r)}$, the evaluation of logical operators will change if we apply the logical rotation and the CNOT gate described in the main text (see Supplementary Information Section I for theoretical details).

\section{Logical CNOT gate and rotation gates in the Floquet-Bacon-Shor code}

Here we provide the theoretical derivation of the logical CNOT gate and the logical rotation gates implemented in our experiments. See Methods for operator definitions.

\subsection{Construction of the logical CNOT gate}

\begin{table}[htbp]
    \centering
    \begin{tabular}{|c|c|c|c||c|c|c|}
        \hline 
        \multicolumn{4}{|c||}{Before $\overline{\rm CNOT}_{\rm L}$} & \multicolumn{3}{c|}{After $\overline{\rm CNOT}_{\rm L}$} \\
        \hline 
        $Z_1$ & $Z_4$ & $Z_7$ & $Z_1Z_4Z_7$ & $Z_1$ & $Z_4$ & $Z_7$ \\
        \hline 
        $+1$ & $+1$ & $+1$ & $+1$ & $+1$ & $+1$ & $+1$ \\
        $+1$ & $+1$ & $-1$ & $-1$ & $-1$ & $+1$ & $+1$ \\
        $+1$ & $-1$ & $+1$ & $-1$ & $-1$ & $-1$ & $-1$ \\
        $+1$ & $-1$ & $-1$ & $+1$ & $+1$ & $-1$ & $-1$ \\
        $-1$ & $+1$ & $+1$ & $-1$ & $+1$ & $+1$ & $-1$ \\
        $-1$ & $+1$ & $-1$ & $+1$ & $-1$ & $+1$ & $-1$ \\
        $-1$ & $-1$ & $+1$ & $+1$ & $-1$ & $-1$ & $+1$ \\
        $-1$ & $-1$ & $-1$ & $-1$ & $+1$ & $-1$ & $+1$ \\
        \hline 
    \end{tabular}    
    \caption{Truth table for $\overline{\rm CNOT}_{\rm L}$. }
    \label{tab:cnot_transform}
\end{table}

At round $i$, the logical CNOT gate between the static and dynamical qubits is defined as 
\[
    {\rm CNOT}_{\rm L} := \ket{0}\bra{0}_{\rm s}\otimes I_{\rm d}+\ket{1}\bra{1}_{\rm s}\otimes X_{\rm d}^{(i)}.
\]
By the equality
\begin{align*}
    {\rm CNOT}_{\rm L} =  \ket{0}\bra{0}_{\rm s}\otimes I_{\rm d}+\ket{1}\bra{1}_{\rm s}\otimes \left(\Gamma_{X}^{(i)}X_{\rm d}\right) 
    = \left(Z_{\rm s}^{\frac{1-\Gamma_{X}^{(i)}}{2}}\otimes I_{\rm d}\right) \left(\ket{0}\bra{0}_{\rm s}\otimes I_{\rm d}+\ket{1}\bra{1}_{\rm s}\otimes X_{\rm d}\right),
\end{align*}
the implementation of ${\rm CNOT}_{\rm L}$ can be decomposed into two parts: (i) $\overline{\rm CNOT}_{\rm L}:=\ket{0}\bra{0}_{\rm s}\otimes I_{\rm d}+\ket{1}\bra{1}_{\rm s}\otimes X_{\rm d}$, which can be viewed as using the value of $Z_{\rm s}$ to control $X_{\rm d}$, and (ii) $Z_{s}$ gate if $\Gamma_{X}^{(i)}=-1$. We first describe the construction of the nFT circuit for part (i). In our experiment, we implement ${\rm CNOT}_{\rm L}$ between stabilizers $S_Z^{\rm B}$ and $S_X^{\rm C}$, where $X_{\rm d}=X_1X_7$. To reduce the circuit depth, we use $Z_1Z_4Z_7$ as $Z_{\rm s}$ instead of $Z_2Z_5Z_8$. Note that for logical states, operators $Z_1Z_4Z_7$ and $Z_2Z_5Z_8$ are equivalent because $Z_1Z_4Z_7=S_Z^{\rm D}\cdot Z_2Z_5Z_8$ and $S_Z^{\rm D}$ does not change logical states. The truth table for $\overline{\rm CNOT}_{\rm L}$ is shown in Table~\ref{tab:cnot_transform}.
From the truth table, we see that the logical CNOT gate can be implemented by applying physical ${\rm CNOT}_{4, 1}$ and ${\rm CNOT}_{4, 7}$, and then swapping $D_1$ and $D_7$. Using standard techniques of quantum circuit construction, we obtain the logical CNOT circuit implemented in our experiment. Lastly, part (ii) is conditioned on previous measurement outcomes, which requires real-time feedback control. However, it can be replaced by data post-processing, that is, multiplying the logical $X$ and $Y$ measurement outcomes of the static qubit with $\Gamma_{X}^{(i)}$, because $Z_{\rm s}$ will flip the logical $X$ and $Y$ measurement outcomes.

\subsection{Construction of the logical rotation gates} 

At round $i$, the logical $Z$-axis rotation $R_{Z}^{\theta}$ and $X$-axis rotation $R_{X}^{\varphi}$ of the dynamical qubit are defined as:
\begin{equation} \label{eq:rotation}
    \begin{split}
    R_{Z}^{\theta}
    &:=\exp\left(-i\frac{\theta}{2} Z_{\rm d}^{(i)}\right)
    =\exp\left(-i\frac{\Gamma^{(i)}_Z\theta}{2} Z_{\rm d}\right),\\
    R_{X}^{\varphi}
    &:=\exp\left(-i\frac{\varphi}{2} X_{\rm d}^{(i)}\right)
    =\exp\left(-i\frac{\Gamma^{(i)}_X\varphi}{2} X_{\rm d}\right).
    \end{split}
\end{equation}
The logical rotation gates can be implemented non-fault-tolerantly through the following three steps: (i) Compute the value of \( Z_{\rm d} \) (\( X_{\rm d} \)) on an ancilla qubit by applying a physical CNOT (\( {\rm CNOT} \cdot (H \otimes I) \)) gate between each involved data qubit and the ancilla qubit. (ii) Apply the physical rotation gate \( R_{z}^{\Gamma^{(i)}_Z\theta} \) (\( R_{x}^{\Gamma^{(i)}_X\varphi} \)). (iii) Apply the inverse of the circuit used in the first step to disentangle the ancilla qubit.


The implementation of $R_{Z}^{\theta}$ is more straightforward between stabilizers $S_Z^{\rm B}$ and $S_X^{\rm C}$, because the logical $Z$ operator after stabilizer $S_Z^{\rm B}$ is $Z_{\rm d}^{(i)}=\Gamma_{Z}^{(i)}Z_7Z_8$, and the involved data qubits $D_7$ and $D_8$ are physically closer on the square lattice. Similarly, $R_{X}^{\varphi}$ is more easier implemented between stabilizers $S_X^{\rm C}$ and $S_Z^{\rm D}$. 

In general, the coefficient $\Gamma^{(i)}_Z$ in the rotation phase $\Gamma^{(i)}_Z\theta$ depends on the outcomes of previous mid-circuit measurements, which typically requires real-time feedback control during the execution of the circuit. In our experiment, the logical $R_{Z}^{\theta}$ gate is applied to $\ket{+}_{\rm d}$, and the real-time feedback control can be replaced by data post-processing, as elaborated below. Using the equality
\[
    \exp\left(-i\frac{\Gamma^{(i)}_Z\theta}{2} Z_{\rm d}\right)\ket{+}_{\rm d}=
    \frac{e^{-\frac{\Gamma^{(i)}_Z\theta}{2}}\ket{0}_{\rm d}+e^{\frac{\Gamma^{(i)}_Z\theta}{2}}\ket{1}_{\rm d}}{\sqrt{2}}=
    X^{\frac{1-\Gamma^{(i)}_Z}{2}}\exp\left(-i\frac{\theta}{2} Z_{\rm d}\right)\ket{+}_{\rm d},
\]
the implementation of $R_{Z}^{\theta}$ gate on $\ket{+}_{\rm d}$ can be decomposed into two parts: (i) A logical rotation $\exp\left(-i\frac{\theta}{2} Z_{\rm d}\right)$ that is independent of mid-circuit measurement outcomes, and (ii) A logical $X$ gate if $\Gamma^{(i)}_Z=-1$. Part (ii) can be implemented by data post-processing: since a logical $X$ gate flips the logical $Z$ and $Y$ measurement outcomes, we multiply the measurement outcomes by $\Gamma^{(i)}_Z$.
Similarly, when applying the logical $R_{X}^{\varphi}$ gate to state $\ket{0}_{\rm d}$, we multiply the logical $X$ and $Y$ measurement outcomes by $\Gamma^{(i)}_X$ during data post-processing.

Lastly, achieving universal control of the two-qubit logical state also requires implementing logical rotation gates on the static qubit. By replacing $Z_{\rm d}^{(i)}$ ($X_{\rm d}^{(i)}$) with $Z_{\rm s}$ ($X_{\rm s}$) in Eq.~\eqref{eq:rotation}, the static logical $R_{Z}^{\theta}$ ($R_{X}^{\varphi}$) gate can be non-fault-tolerantly implemented in the same manner as its dynamical counterpart. However, unlike in the dynamical case, the data post-processing is not required, as the state operators remain unchanged during the stabilizer measurement.

\section{Two-qubit Logical Quantum State/Process Tomography}

\subsection{Two-qubit Logical Quantum State Tomography}
The construction of the logical density matrix $\rho_{\rm L}$ in the \(Z\) basis follows a procedure similar to that used for a physical density matrix. For a physical density matrix, single-qubit gates \(\{I,R_y^{\pi/2},R_x^{\pi/2}\}\) are applied to the physical qubits before performing projective measurements, enabling measurements in the \(X\), \(Y\), and \(Z\) directions. However, for logical states, these rotation operations are unnecessary. Instead, we directly perform logical measurements in the \(X\), \(Y\), and \(Z\) bases.
For a two-qubit logical state, we measure the state in 9 logical bases $\{Z,X,Y\}^{\otimes 2}$. For each basis, we obtain an array $[p_{--},p_{-+},p_{+-},p_{++}]$, where $p_{-+}$ denotes the probability of measuring the static qubit as $-1$ and the dynamical qubit as $+1$.

We optimize the logical density matrix $\rho_{\rm L}$ using the convex optimization package cvxpy~\cite{diamond2016cvxpy}, constraining the density matrix to be Hermitian, unit-trace and positive semi-definite.
The fidelity of the reconstructed state relative to a target pure state is calculated as ${\rm Tr}(\rho_{\rm L} \rho_{\rm ideal})$.

\subsection{Two-qubit Logical Quantum Process tomography}

LQPT is built upon the results of LQST.
Each two-qubit logical density matrix $\rho_{\rm L}$ can be expressed in the two-qubit Pauli basis as a vector $\vec{p}^{\mathrm{T}} = [p_{00},p_{01},...,p_{32},p_{33}]$, where
\begin{equation*}
     p_{ij}= \frac{{\rm Tr}(\rho_{\rm L} \sigma_i \otimes \sigma_j )}{ {\rm Tr}(\rho_{\rm L}) },\text{and} ~ \sigma_i,\sigma_j \in \{I,X,Y,Z\}.
\end{equation*}
The LPTM $R$, a $16\times16$ matrix, maps an input state vector $\vec{p}$ to an output state vector $\vec{p'} = R \vec{p}$~\cite{chow2012universal}.
We reconstruct the LPTM using a complete set of input-output state vectors.
Convex optimization is employed to find an optimal LPTM that satisfies all the constraints of a physical quantum channel.

\section{Experimental setup and device parameters}

The experiments described in this manuscript are conducted on a superconducting quantum processor comprising 66 qubits and 110 tunable couplers. For details on the chip structure, fabrication process, and packaging scheme, refer to~\cite{yang2024coupler} and~\cite{huang2025exact}. A subset of the processor comprising 21 qubits and 32 couplers is used in this experiment. The qubit parameters, coherence properties, and gate performance of the subset are summarized in Table~\ref{tabel_qubits}.

\begin{figure}[htbp]
    \centering
    \includegraphics[width=\textwidth]{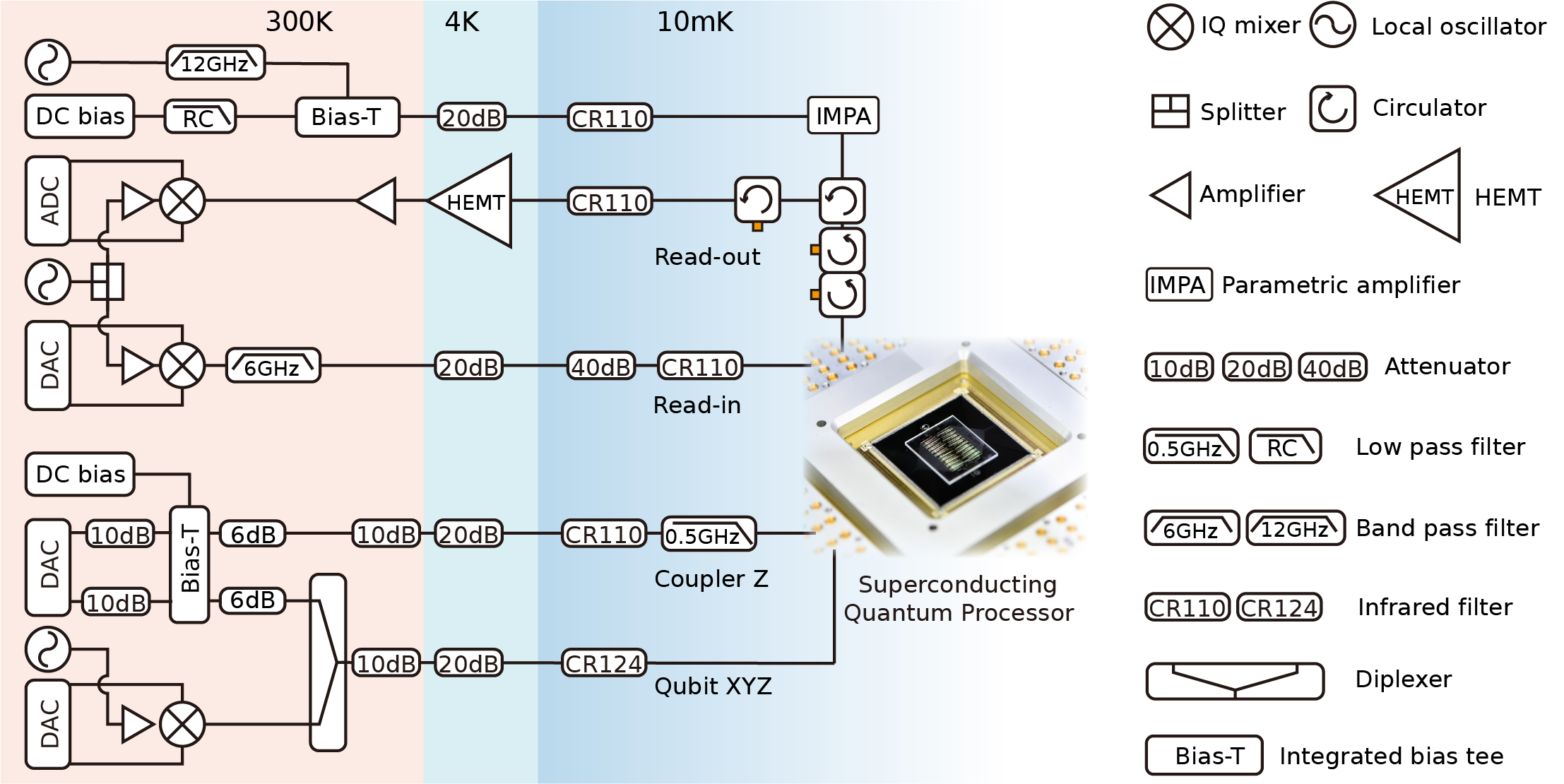}
    \caption{Schematic of the experimental setup.} \label{fig:wiring}
\end{figure}

The quantum processor is mounted on the mixing chamber plate of a dilution refrigerator, operating at a temperature of approximately 10~mK. Detailed diagrams of the control electronics, wiring, and filtering are shown in Fig.~\ref{fig:wiring}. 
Microwave measurement and control system (M2CS)~\cite{Zhang2024} are used for generating control signals and acquiring data, including digital-to-analog converter (DAC) and analog-to-digital converter (ADC) circuit boards. 
Integrated bias-tees are employed to combine a shared DC bias signal with qubit/coupler flux pulses. 
This configuration enables the use of additional attenuators at the DAC ports, which helps suppress low-frequency noise and thereby enhances the Ramsey dephasing time.
In the readout line, impedance-matched parametric amplifiers (IMPAs)~\cite{mutus2014strong} are utilized for the initial stage of signal amplification, followed by high electron mobility transistor (HEMT) amplifiers and room-temperature amplifiers. All control signals are routed through multiple stages of attenuation and filtering to prevent noise from entering the quantum processor.

\begin{table}
\begin{ruledtabular}
\begin{tabular}{lccc}
Parameters & Mean & Median & Stdev. \\
\hline
Qubit idle frequency (GHz) & 4.195 & 4.223 & 0.143 \\
Qubit anharmonicity (MHz) & $-214.2$ & $-215.0$ & 6.1 \\
Readout frequency (GHz) & 6.213 & 6.218 & 0.063 \\
$T_1$ at idle frequency ($\mu$s) & 77.2 & 72.7 & 17.8 \\
$T_2^{*}$ at idle frequency ($\mu$s) & 3.87 & 3.85 & 0.71 \\
$T_2^{\rm echo}$ at idle frequency ($\mu$s) & 11.74 & 11.50 & 2.85 \\
QND Readout error (\%, simultaneous) & 1.63 & 1.58 & 0.61 \\
Single qubit gate error ({\textperthousand}, simultaneous) & 0.76 & 0.71 & 0.24 \\
CZ gate error (\%, simultaneous) & 1.01 & 0.97 & 0.30 \\
Idle error with DD (\%, simultaneous) & 1.46 & 1.43 & 0.39 \\
\end{tabular}
\end{ruledtabular}
\caption{\label{tabel_qubits} Parameters of the superconducting quantum processor.}
\end{table}

\section{Readout optimization of the ancilla qubits}

\subsection{Quantifying readout error using repeated measurements}

\begin{figure}[htbp]
    \centering
    \includegraphics[width=0.9\textwidth]{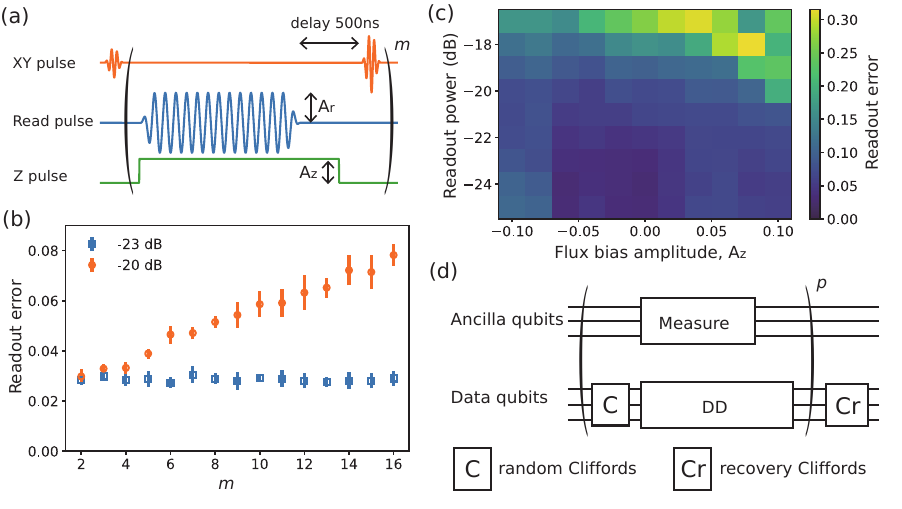}
    \caption{Optimizing readout parameters and benchmarking data qubit errors during ancilla readout.
    (a) Experimental sequence for repeated measurements to determine readout errors.
    The qubit is initialized in a superposition state using an $R_x^{\pi/2}$ pulse before the repeated pulses. A 500~ns delay is inserted between the readout pulse and the $R_x^{\pi}$ pulse to deplete residual photons in the resonator. The readout pulse amplitude and frequency, as well as the qubit flux bias amplitude, are optimized to minimize readout error.
    (b) Readout error as a function of the number of repeated measurement cycles $m$ for different (relative) readout powers.
    (c) Readout error as a function of the flux bias amplitude $A_{\rm z}$ and the readout pulse amplitude (converted to power units), at a fixed number of repeated measurement cycles $m=10$.
    (d) Circuit for the interleaved randomized benchmarking experiment, which measures idle errors of data qubits (with dynamical decoupling pulses) during ancilla qubit readout.
    } \label{fig:qnd_read}
\end{figure}

To quantify the quantum non-demolition (QND) readout error of an ancilla qubit, we repeatedly perform a composite operation consisting of a readout pulse followed by an $R_x^{\pi}$ pulse, with the qubit initialized in a superposition state. 
The experimental pulse sequence is shown in Fig.~\ref{fig:qnd_read}(a). The readout pulse duration is fixed at 500~ns, and a 500~ns delay for resonator photon depletion is inserted between the readout pulse and the $R_x^{\pi}$ pulse.
The readout error is defined as $\varepsilon_r = 1 - P(0|1)/2 - P(1|0)/2$, where \(P(0|1)\) is the probability of measuring the qubit as 0 given that the previous measurement result was 1, and \(P(1|0)\) denotes the opposite scenario.

The signal-to-noise ratio (SNR) of the measurement signal is directly related to the readout pulse amplitude \(A_{\rm r}\). Increasing \(A_{\rm r}\) reduces the separation error $\varepsilon_{\rm sep} = \frac{1}{2} {\rm erfc}(\frac{\sqrt{\rm SNR}}{2})$. For a given readout pulse, we optimize the demodulation window to maximize the SNR~\cite{bengtsson2024model}. However, higher readout power may cause state leakage out of the computational subspace~\cite{sank2016measurement}, leading to an increase in readout error with the number of repeated measurement cycles $m$, as shown in Fig.~\ref{fig:qnd_read}(b). Additionally, higher readout power can induce more cross-measurement dephasing on adjacent data qubits~\cite{bultink2020protecting}. 
Since the measurement-induced state transition rate is strongly related to the resonator photon number during readout~\cite{hazra2024benchmarking}, we set the readout pulse frequency to be centered between the two dressed resonator states, This ensures that the resonator photon number is similar for both qubit states, and the Stark shifts caused by photons in the resonator are also similar.
The dressed resonator frequencies are determined using a kappa-chi-power experiment~\cite{sank2025system}.
The qubit frequency during readout is adjusted by the flux pulse amplitude $A_{\rm z}$ to avoid two-level system defects, which can cause significant relaxation errors~\cite{thorbeck2024readout}. We optimize the readout error at a fixed number of repeated measurement cycles $m=10$ by scanning the readout pulse amplitude and the qubit frequency during readout, as shown in Fig.~\ref{fig:qnd_read}(c).

\subsection{Data qubit idle error during ancilla readout}

The readout pulses and the idle time for resonator photon depletion account for 720~ns, which is approximately 78\% of the time spent in a stabilizer round. During the period, data qubits are subject to energy relaxation and dephasing errors. Additionally, the ancilla qubit measurement process introduces extra phase errors in the data qubits, primarily due to photon crosstalk between readout resonators~\cite{szombati2020quantum}. 
To mitigate these dephasing errors, we employ a dynamical decoupling (DD) technique by inserting a Carr-Purcell-Meiboom-Gill (CPMG) sequence on the data qubits during the 720~ns measurement time. The sequence consists of eight $R_y^{\pi}$ gates.

We assess the idle errors of the data qubits during ancilla readout using interleaved randomized benchmarking (RB)~\cite{magesan2012efficient}. The measurement sequence, which includes simultaneous readout pulses on all ancilla qubits and DD sequences on all data qubits, is inserted between single-qubit Clifford gates on the data qubits. The RB circuit is shown in Fig.~\ref{fig:qnd_read}(d). 
The idle error of the data qubits is sensitive to the ancilla qubit frequency during measurement. To minimize idle errors caused by stray interactions, the ancilla qubit frequency during readout (including Stark shift induced by resonator photons) is adjusted by the flux pulse to avoid resonances between the $|10\rangle$ and $|01\rangle$, as well as between $|11\rangle-|20\rangle$ ($|11\rangle-|02\rangle$).

\section{Leakage accumulation during repeated stabilizer measurements}

\begin{figure}[htbp]
    \centering
    \includegraphics[width=0.5\textwidth]{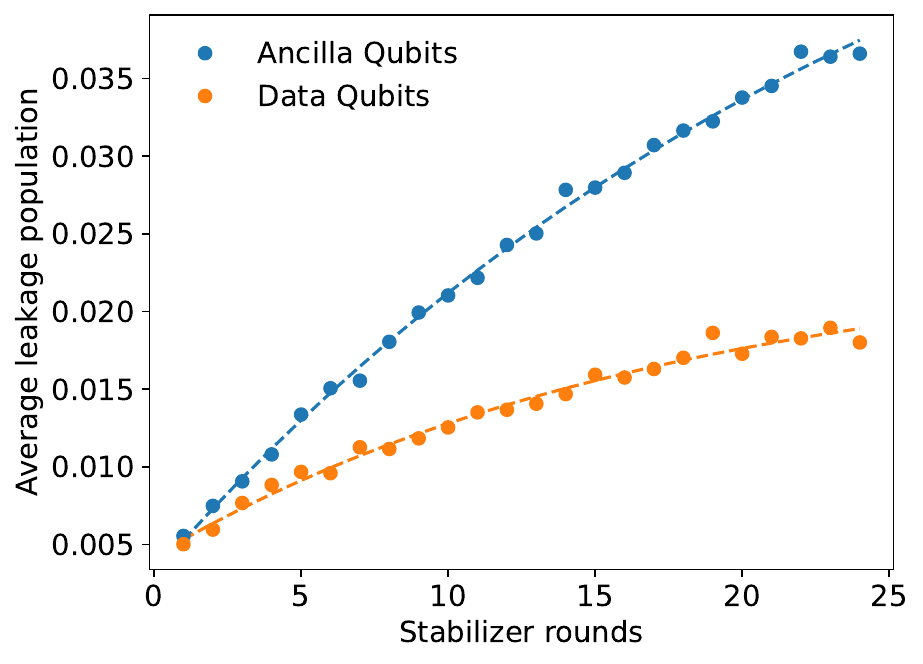}
    \caption{Average leakage accumulation on the ancilla (blue) and data (orange) qubits as a function of stabilizer rounds. The dashed lines indicate exponential fits. The fitted average leakage rates are 0.22\% per round for ancilla qubits and 0.14\% per round for data qubits.
    } \label{fig:leakage}
\end{figure}

We investigate the leakage into the second excited states of the qubits during the logical state preservation experiments by performing three-state readout on all qubits at the end of multi-round stabilizer measurements. 
The average leakage populations versus stabilizer rounds for the ancilla and data qubits are shown in Fig.~\ref{fig:leakage}. 
The leakage population $p_{\rm leak}$ increases with the number of stabilizer rounds $r$, following the relationship: $ {\rm d} p_{\rm leak} / {\rm d} r = \epsilon_{\rm leak} - p_{\rm leak} \cdot \epsilon_{\rm d}$, where $\epsilon_{\rm leak}$ is the leakage rate per round, and $\epsilon_{\rm d}$ is the average decay rate per round of the second excited states. 
We fit the leakage data using the function $p_{\rm leak} = \epsilon_{\rm leak}/\epsilon_{\rm d} - (\epsilon_{\rm leak}/\epsilon_{\rm d} - p_0) \cdot e^{-\epsilon_{\rm d} r}$, where $p_0$ is the initial leakage population at round 0. The fitted leakage rates are 0.14\% per round for the data qubits and 0.22\% per round for the ancilla qubits.
The primary source of leakage is attributed to the CZ gates, as their implementation involves transitions to the second excited states~\cite{sung2021realization}. 
The higher leakage rate observed for the ancilla qubits is due to the greater involvement of their second excited states in the implementation of the CZ gates.

\section{Bacon-Shor code}\label{sec2}

Here we present experimental results for the distance-3 BS code, using the same qubits as in the FBS code. We demonstrate logical state encoding and measurement (Fig.~\ref{fig:exp01_BS_encode}), multi-round stabilizer measurements (Fig.~\ref{fig:exp02_BS_memory} and Fig.~\ref{fig:exp02_BS_memory_result}) and FT logical gates (Fig.~\ref{fig:exp03_BS_sqg}).

\subsection{Logical state encoding and measurement}

\begin{figure}[htbp]
    \centering
    \includegraphics[width=0.9\textwidth]{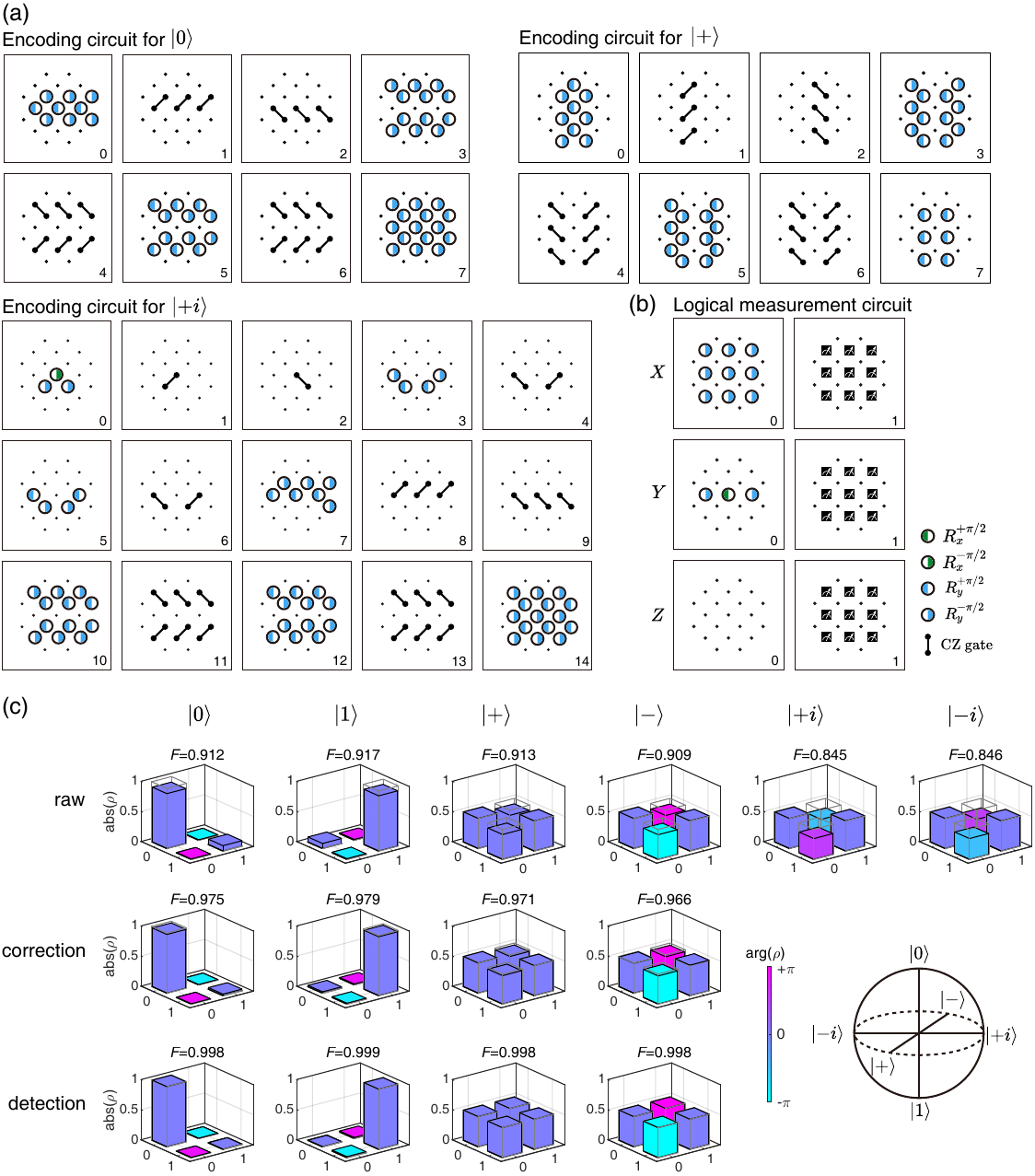}
    \caption{Encoding logical states in the BS code.
    (a) Encoding circuits for logical states $|0\rangle$, $|+\rangle$, and $|{+}i\rangle$, with the gate patterns arranged in order (numbered at the bottom-right corner).
    (b) Logical measurement circuits for the BS code: $X$ (top), $Y$ (middle), and $Z$ (bottom).
    (c) LQST results for encoded Pauli states: without post-processing (top), with error correction (middle), and with error detection (bottom). The wireframes represent the amplitudes of the ideal density matrices.
    } \label{fig:exp01_BS_encode}
\end{figure}

Logical states $\{|+\rangle,|-\rangle,|0\rangle,|1\rangle\}$ are fault-tolerantly prepared because errors cannot propagate among data qubits arranged in different rows (for $|0\rangle$ and $|1\rangle$) or columns (for $|+\rangle$ and $|-\rangle$). The encoding circuits for $|0\rangle$ and $|+\rangle$ are shown in Fig.~\ref{fig:exp01_BS_encode}(a). This fault-tolerance does not extend to the state $|{+}i\rangle$, which requires a more complex encoding circuit. The encoding circuits for $|1\rangle$, $|-\rangle$, and $|{-}i\rangle$ differ from their counterparts by the addition of logical Pauli gate circuits, as discussed later.

The logical measurement circuit is shown in Fig.~\ref{fig:exp01_BS_encode}(b). All data qubits are measured simultaneously, enabling final round error detection. The $X$ and $Z$ measurements are FT because the corresponding stabilizers are equivalently performed to detect any single qubit error.

We show the fidelities of the encoded states in Fig.~\ref{fig:exp01_BS_encode}(c), with error correction and error detection applied to states $\{|+\rangle,|-\rangle,|0\rangle,|1\rangle\}$.
The state fidelities are benchmarked using single-qubit LQST. The near-unity fidelities achieved with error detection indicate that most errors has been successfully detected during the logical measurement. 
In contrast, error detection or error correction are not performed for the states $|\pm i\rangle$ due to the nFT nature of the logical $Y$ measurement.

\subsection{Logical state preservation}

\begin{figure}[htbp]
    \centering
    \includegraphics[width=\textwidth]{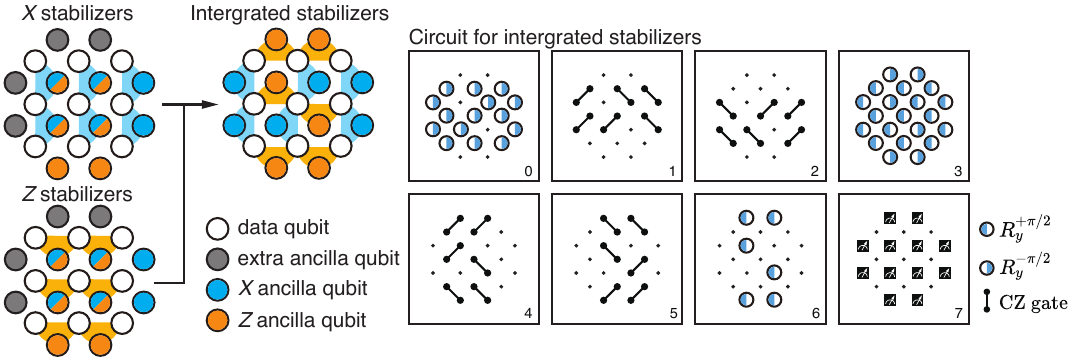}
    \caption{
    Integrating $X$ and $Z$ stabilizers in the BS code.
    We introduce additional ancilla qubits (grey circles) on the left and top of the lattice to combine the $X$ and $Z$ stabilizers into a single stabilizer round for the distance-3 BS code, as indicated by the arrow. Each stabilizer round now includes four weight-6 stabilizers: two $X$ stabilizers and two $Z$ stabilizers. The detailed circuit for the integrated stabilizer is shown on the right.
    } \label{fig:exp02_BS_memory}
\end{figure}

\begin{figure}[htbp]
    \centering
    \includegraphics[width=0.9\textwidth]{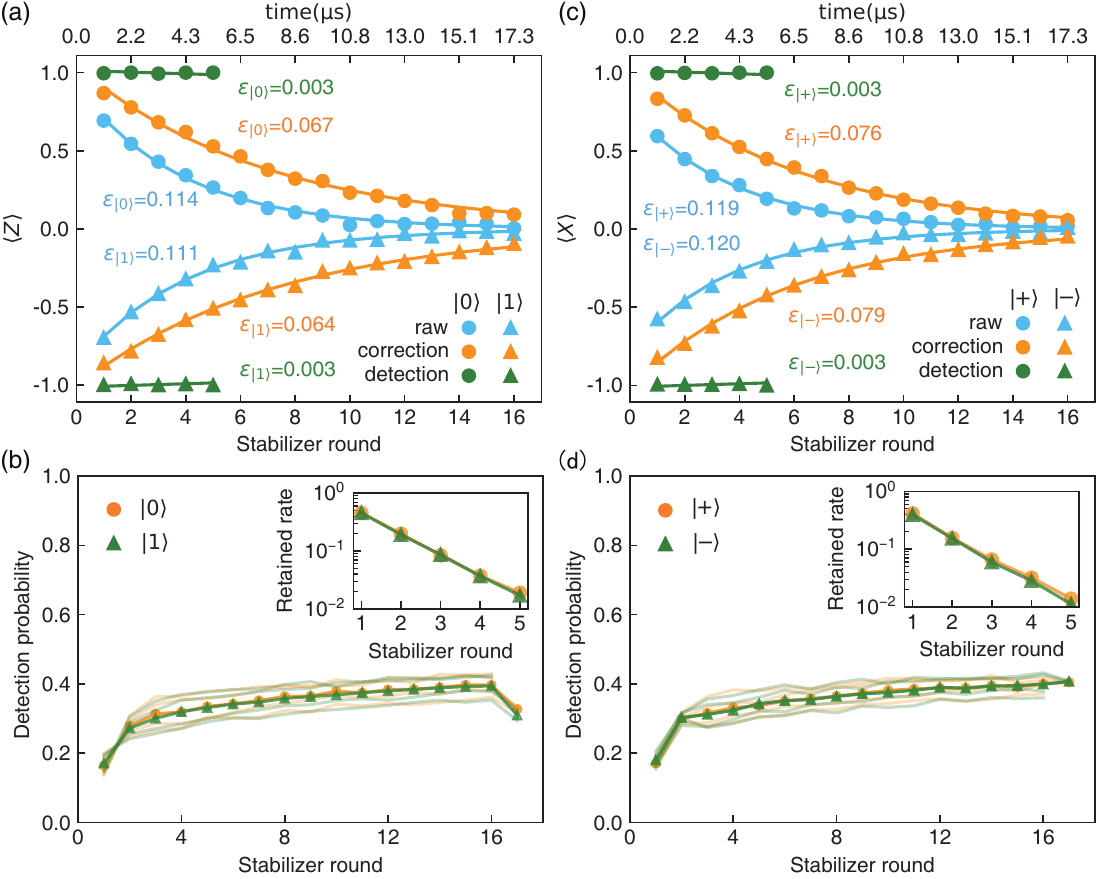}
    \caption{
    Logical state preservation in the BS code with integrated $X$ and $Z$ stabilizer measurements.
    (a) Measured expectation values of the logical $Z$ operator versus stabilizer round for prepared $|0\rangle$ (circles) and $|1\rangle$ (triangles). 
    The cyan, green, and orange colors correspond to data without post-processing, data with error detection, and data with error correction, respectively. Solid lines represent exponential fits.
    (b) Detection probability for each weight-6 stabilizer over 16 rounds of stabilizer measurements, for encoded states $|0\rangle$ (orange) and $|1\rangle$ (green). 
    Darker lines show the average detection probabilities for the two encoded states. Detection probabilities are lower in the first and last rounds, where data qubit initialization and measurement values are used to identify parity errors.
    A slight increase in detection probability is observed in the intermediate rounds, likely due to accumulated leakage errors. The inset shows the exponential decrease in the retained data rate during error detection, based on 10,000 experimental shots, with corresponding exponential fits (solid lines).
    (c) and (d), Same as (a) and (b), but for logical states  $|+\rangle$  and  $|-\rangle$ .
    } \label{fig:exp02_BS_memory_result}
\end{figure}

In the conventional BS code, $X$ and $Z$ stabilizer measurements are performed alternatively in a 2D grid, with the four central qubits alternatively used as $X$ and $Z$ ancilla qubits. 
Here, we merge the $X$ and $Z$ stabilizers into a single stabilizer round using extra ancilla qubits. 
Consequently, each stabilizer round consists of four weight-6 stabilizers (two $X$ stabilizers and two $Z$ stabilizers).
The corresponding circuit is shown in Fig.~\ref{fig:exp02_BS_memory}.
We present the results of logical state preservation for encoded states $\{|+\rangle,|-\rangle,|0\rangle,|1\rangle\}$ in Fig.~\ref{fig:exp02_BS_memory_result}.
An error detection event occurs when any stabilizer measurement result deviates from its theoretical value. Stabilizer measurement results are compared to those from the previous round, except for the first round. A lower detection probability is observed in the first (last) round because data qubit initialization (measurement) values are used to identify parity errors and generate detection events. A slight increase in detection probability is observed in the intermediate rounds, likely due to the accumulation of leakage errors. 

Notably, we do not apply reset operations to the ancilla qubits between stabilizer rounds. Instead, we perform XOR operations on the measurement results of the ancilla qubits during post-processing~\cite{kelly2015state}. Due to the exponential drop in the retained rate for error detection, we limit our analysis to five rounds, with each experiment comprising 10,000 measurement shots.

\subsection{Logical gates}

\begin{figure}[htbp]
    \centering
    \includegraphics[width=0.9\textwidth]{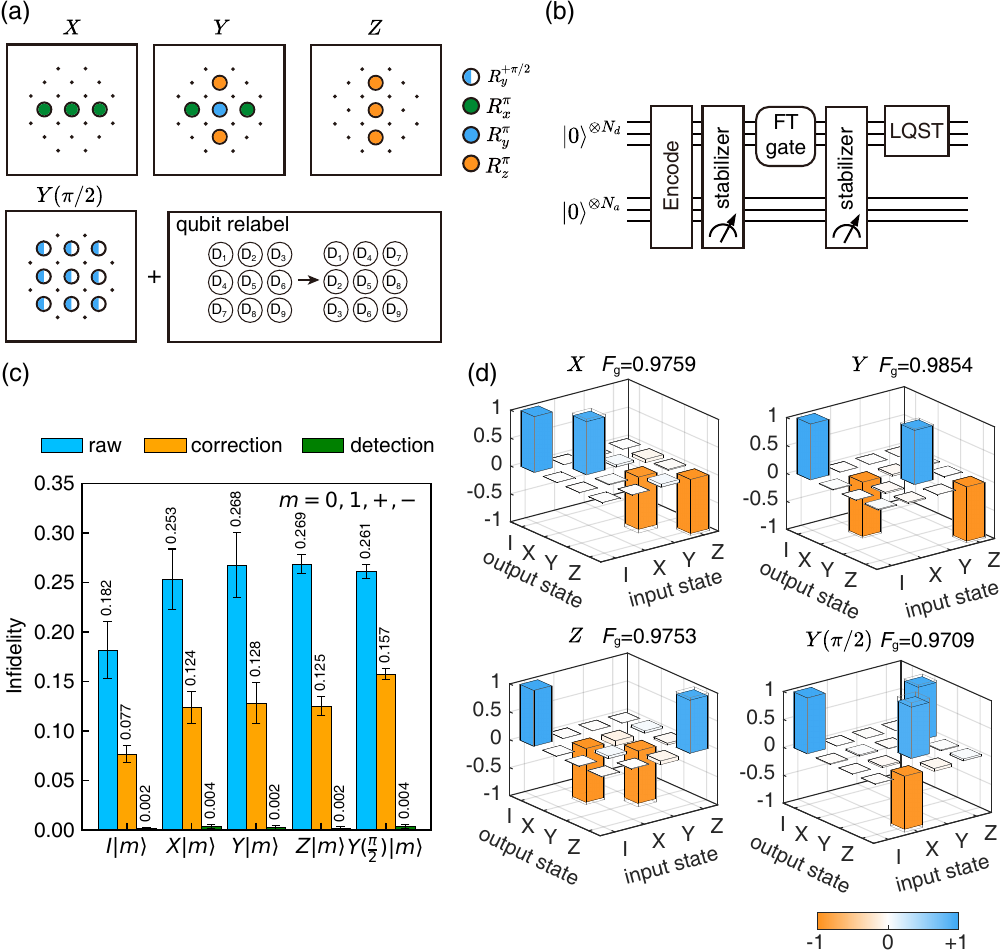}
    \caption{
    Logical gates in the BS code.
    (a) Circuits for the logical $X$, $Y$, $Z$ and $Y(\pi/2)$ gate. The $Y(\pi/2)$ gate is implemented by applying physical $R_{y}^{\pi/2}$ gates on all the data qubits and relabeling the data qubit indices in post-processing.
    (b) Circuit for implementing the FT logical gates in the BS code. The transversal gates on the data qubits are inserted between two integrated stabilizers, and LQST is used to determine the final state fidelity. 
    (c) Average state infidelities over four encoded states after applying the respective logical gates and two stabilizer rounds, without post-processing (blue), with error detection (green) and with error correction (orange). Error bars correspond to 95\% confidence intervals. For the identity gate, no gate circuit is inserted, and only one stabilizer round is applied after the encoding circuit. 
    (d) Constructed LPTMs for the $X$, $Y$, $Z$ and $Y(\pi/2)$ gates, with error detection applied to both the input states and output states.
    Both the input and output states are characterized using LQST. 
    } \label{fig:exp03_BS_sqg}
\end{figure}

The FT logical gates are implemented transversally by applying parallel single qubit rotation gates on relevant data qubits. 
The gate circuits for the logical $X$, $Y$, $Z$, and $Y(\pi/2)$ gates are shown in Fig.~\ref{fig:exp03_BS_sqg}(a). 
The $Y(\pi/2)$ gate is implemented by applying $R_{y}^{\pi/2}$ gates on all the data qubits and reassigning the data qubit indices during post-processing.
We insert the transversal gates between two integrated stabilizer rounds and perform LQST on the final states, as shown in Fig.~\ref{fig:exp03_BS_sqg}(b).
The infidelities of the final-states, averaged over 4 encoded states $\{|0\rangle, |1\rangle, |+\rangle, |-\rangle \}$, are compared to the infidelity of the identity gate (i.e., with no gate inserted), as shown in Fig.~\ref{fig:exp03_BS_sqg}(c). 
To assess the performance of the logical gates with error detection, we perform LQPT using six encoded cardinal states $\{|0\rangle, |1\rangle, |+\rangle, |-\rangle, |{+}i\rangle, |{-}i\rangle \}$ as inputs. The LPTMs extracted from the over-complete sets of input-output logical-state pairs are shown for the $X$, $Y$, $Z$, $Y(\pi/2)$ gates in Fig.~\ref{fig:exp03_BS_sqg}(d).

\section{Floquet-Bacon-Shor code}\label{sec2}

\subsection{Logical state encoding and measurement}

\begin{figure}[h!]
    \centering
    \includegraphics[width=\textwidth]{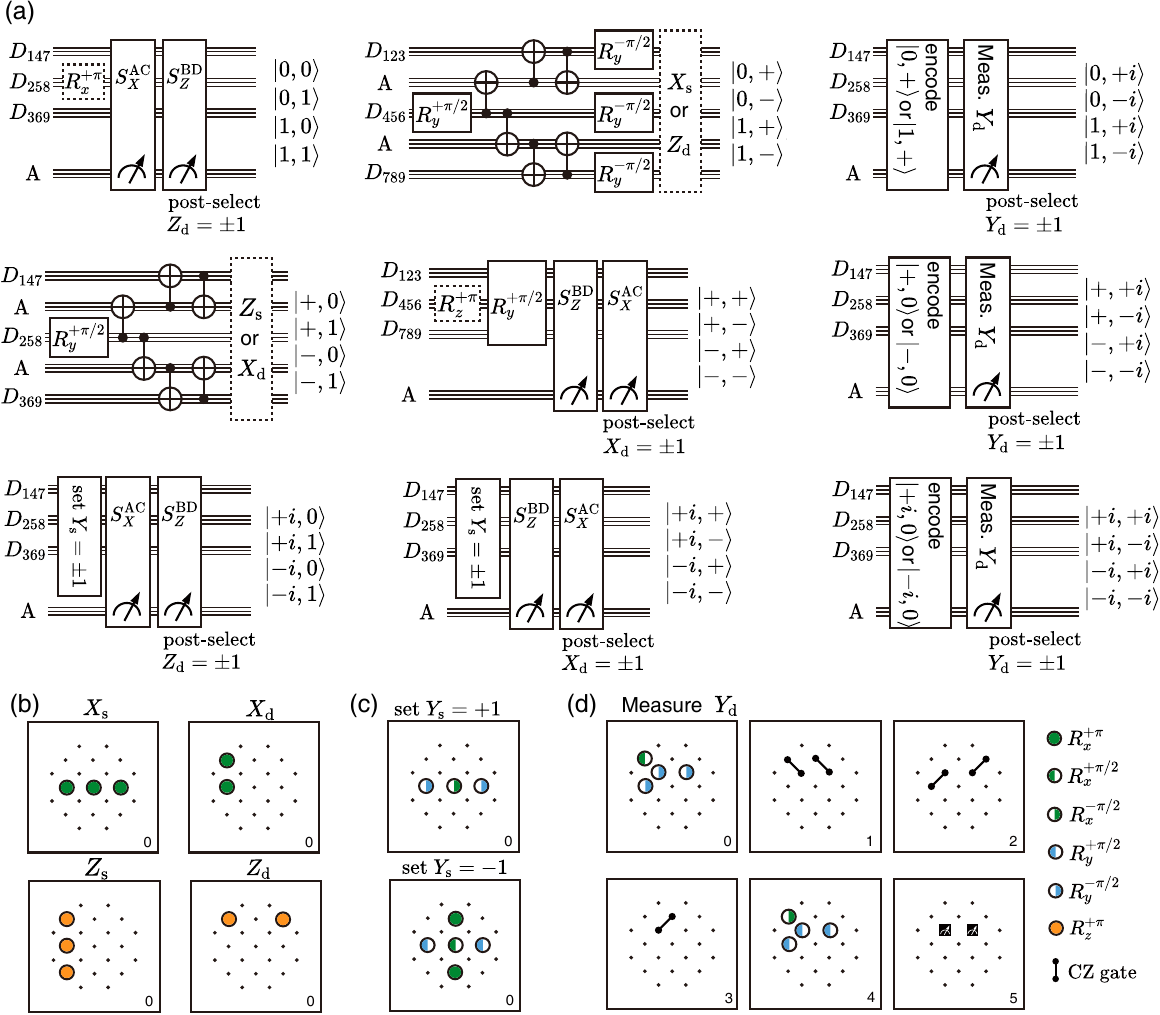}
    \caption{
    Encoding circuits for Pauli states in the FBS code.
    (a) Encoding circuits for 36 Pauli states. The 16 states in the group $\{|
    |0\rangle,|1\rangle,|+\rangle,|-\rangle\}^{\otimes 2}$ are fault-tolerantly prepared.
    The dashed box indicates the gate circuit for FT Pauli gates on the data qubits, detailed in (b).
    The CNOT symbol denotes transversal CNOT gates between data and ancilla qubit
    groups, subject to connectivity constraints. For example, in the $ZX$-basis encoding circuit, the topmost $A$ refers to the group of $3$ ancillas located
    between data-qubit groups $D_{123}$ and $D_{456}$, and the CNOT gates are
    applied transversally between $A$ and $D_{123}$ ($D_{456})$.
    (b) Circuits for logical $X$ (top) and $Z$ (bottom) gates applied to the static (left) and dynamical (right) qubits. 
    (c) Circuits for setting the $Y_{\rm s}$ operator of the static qubit to $+1$ (top) and $-1$ (bottom). 
    (d) Circuit for setting the $Y_{\rm d}$ operator of the dynamical qubit using ancilla measurements. 
    The dynamical qubit is set to $|{+}i\rangle$ ($|{-}i\rangle$) if the measured parity of the two ancilla qubits is $+1$ ($-1$).
    } \label{fig:exp07_FBS_encode}
\end{figure}

\begin{figure}[htbp]
    \centering
    \includegraphics[width=0.6\textwidth]{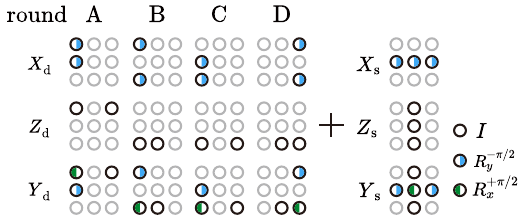}
    \caption{Physical rotation gates applied to the data qubits before readout pulses for measuring logical operators. The rotation gates for logical operators of the dynamical and static qubits are combined to enable simultaneous logical measurements. For the dynamical qubit, the specific rotation gates applied vary according to the type of preceding stabilizer.
    } \label{fig:FBS_measure}
\end{figure}

\begin{figure}[htbp]
    \centering
    \includegraphics[width=0.6\textwidth]{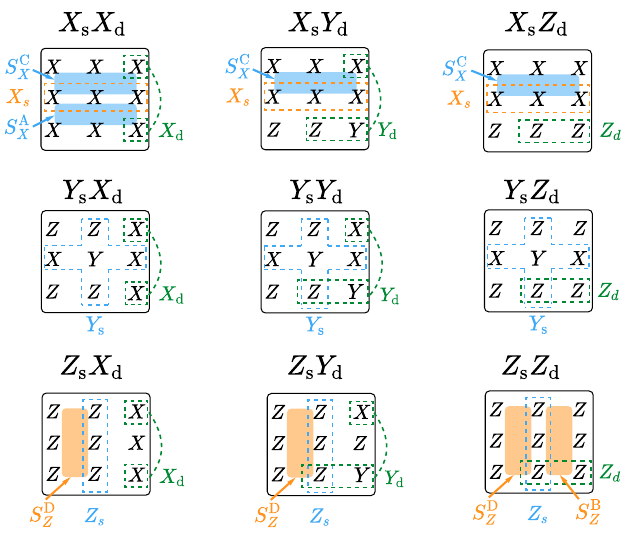}
    \caption{Two-qubit logical measurement circuits in 9 bases after stabilizer $S_X^{\rm D}$. The $X_{\rm s}X_{\rm d}$ and $Z_{\rm s}Z_{\rm d}$ operators are fault-tolerantly measured, with two stabilizer values are calculated to detect any single-qubit error. Although measurements of $X_{\rm s}Y_{\rm d}$, $X_{\rm s}Z_{\rm d}$, $Z_{\rm s}X_{\rm d}$, and $Z_{\rm s}Y_{\rm d}$ are not FT, some errors can still be detected by a single stabilizer.
    } \label{fig:FBS_measure_all}
\end{figure}

We present the encoding circuits for 36 Pauli states in Fig.~\ref{fig:exp07_FBS_encode}(a). The states in the subgroup $\{|0\rangle,|1\rangle,|+\rangle,|-\rangle \}^{\otimes2}$ are fault-tolerantly prepared. Pauli states with the static and dynamical qubits in different bases, as $\{\ket{0,+/-}, \ket{1,+/-}, \ket{+,0/1}$, $\ket{-,0/1}\}$, are products of GHZ states. The state $\ket{+,0}$ differs from $\ket{0,+}$ by entangling data qubits in rows or columns. In the latter case, physical $R_y^{-\pi/2}$ gates are applied to all data qubits after the entangling circuit. 
The remaining states in the group are encoded by applying additional logical $X$ or $Z$ gates to either the static or dynamical qubit. The corresponding gate circuits are shown in Fig.~\ref{fig:exp07_FBS_encode}(b). For Pauli states with both the static and dynamical qubits in the $X$ ($Z$) basis, the state of the dynamical qubit depends on the outcome of the previous $X$ ($Z$) stabilizer measurement.

We analyze the fault tolerance of the encoding circuits within the error-detection framework, where an operation is considered fault-tolerant if any single-qubit error is detectable by stabilizer measurements. For logical states prepared in the $XZ$ ($ZX$) basis, the encoding circuits coincide with those for the $X$ ($Z$) basis states of the BS code. Consequently, the static logical qubit inherits the fault tolerance of the BS code. For the dynamical logical qubit, initialized in the $Z$ ($X$) basis, any single-qubit $Z$ ($X$) error leaves the logical state unchanged. A single-qubit $X$ ($Z$) error may propagate along a row (column), but such propagation is detected by the subsequent $Z$-type ($X$-type) stabilizer measurements—unless it propagates into a static logical $X$ ($Z$) operator, in which case the error stabilizes the state and remains harmless. For logical states in the $XX$ ($ZZ$) basis, the encoding is also fault-tolerant: the initial transversal rotations do not propagate errors and prepare the $X$-type ($Z$-type) stabilizers with deterministic $+1$ outcomes. Any single-qubit $Z$ ($X$) error is detected in the subsequent stabilizer round, while any single-qubit $X$ ($Z$) error does not affect the logical $XX$ ($ZZ$) state.

Two-qubit states involving \(\ket{\pm i}\) are prepared in a nFT manner. The static qubit is encoded into the \(\ket{\pm i}\) state by first setting \(Y_{\rm s} = \pm 1\) (the corresponding circuits are illustrated in Fig.~\ref{fig:exp07_FBS_encode}(c)) and then projecting the system onto the logical subspace via BS stabilizer measurements. 
However, all four stabilizers take on random initial values, preventing their use for error detection. The dynamical qubit is encoded into the \(\ket{\pm i}\) state using \(Y_{\rm d}\)-basis measurements, with logical states post-selected. 
To reduce circuit depth, instead of measuring \( Y_{\rm d} \) directly, we decompose it at round \( A \) into two gauge operators \(Y_{\rm d} = Y_1Z_3X_4 = Y_1X_4Z_2 \cdot Z_2Z_3\) and then measure \( Y_1X_4Z_2 \) and \( Z_2Z_3 \) separately (measurement circuit illustrated in Fig.~\ref{fig:exp07_FBS_encode}(d)). If the product of these two measurement outcomes is \( +1 \) (\(-1\)), the logical state will be projected onto \( \ket{i}_{\rm d} \) (\( \ket{-i}_{\rm d} \)) by the subsequent stabilizer \( S_{X}^{\rm A} \). This process is also nFT due to the inherently nFT nature of the logical $Y$ measurement.

Physical rotation gates are applied to the data qubits before the readout pulses to measure the relative logical operators for the dynamical (left) and static (right) qubits, are shown in Fig.~\ref{fig:FBS_measure}. 
The circuit for the dynamical qubit varies according to the type of preceding stabilizer. The circuits for the static and dynamical logical qubits are combined to perform simultaneous logical measurements. 
We present the measurement circuits in 9 logical bases after the stabilizer \(S_X^{\rm D}\) in Fig.~\ref{fig:FBS_measure_all}. Data qubits that do not participate in defining the logical operators are measured to enable error detection.

The encoded state fidelities of the 36 Pauli states are summarized in Table~\ref{table:fidelity}, with and without error detection during logical measurement. 
The fidelities are benchmarked used LQST. Error detection shows negligible improvement for $\ket{\pm i, \pm i}$ states because no stabilizer values can be calculated from the $Y_{\rm s}Y_{\rm d}$ measurement outcomes.

\begin{table}[h]
\centering
\begin{tabular}{p{1.8cm}p{1.8cm}p{2cm}p{2cm}||p{1.8cm}p{1.8cm}p{2cm}p{2cm}}
\hline
\hline
Encoded state & Fidelity (raw) & Fidelity (detection) & Fault-tolerance & Encoded state & Fidelity (raw) & Fidelity (detection) & Fault-tolerance \\
\hline
$\ket{0,0}$ & 0.775 & 0.898 & FT & $\ket{+i,0}$ & 0.687 & 0.763 & nFT \\
$\ket{0,1}$ & 0.771 & 0.897 & FT & $\ket{+i,1}$ & 0.679 & 0.756 & nFT \\
$\ket{1,0}$ & 0.762 & 0.891 & FT & $\ket{-i,0}$ & 0.703 & 0.767 & nFT \\
$\ket{1,1}$ & 0.770 & 0.896 & FT & $\ket{-i,1}$ & 0.685 & 0.761 & nFT \\
\hline
$\ket{0,+}$ & 0.847 & 0.949 & FT & $\ket{0,+i}$ & 0.696 & 0.831 & nFT \\
$\ket{0,-}$ & 0.832 & 0.949 & FT & $\ket{0,-i}$ & 0.706 & 0.845 & nFT \\
$\ket{1,+}$ & 0.841 & 0.948 & FT & $\ket{1,+i}$ & 0.717 & 0.843 & nFT \\
$\ket{1,-}$ & 0.847 & 0.946 & FT & $\ket{1,-i}$ & 0.730 & 0.848 & nFT \\
\hline
$\ket{+,0}$ & 0.805 & 0.931 & FT & $\ket{+,+i}$ & 0.626 & 0.754 & nFT \\
$\ket{+,1}$ & 0.799 & 0.927 & FT & $\ket{+,-i}$ & 0.611 & 0.715 & nFT \\
$\ket{-,0}$ & 0.786 & 0.923 & FT & $\ket{-,+i}$ & 0.610 & 0.738 & nFT \\
$\ket{-,1}$ & 0.784 & 0.919 & FT & $\ket{-,-i}$ & 0.614 & 0.736 & nFT \\
\hline
$\ket{+,+}$ & 0.752 & 0.878 & FT & $\ket{+i,+}$ & 0.717 & 0.782 & nFT \\
$\ket{+,-}$ & 0.756 & 0.879 & FT & $\ket{+i,-}$ & 0.718 & 0.777 & nFT \\
$\ket{-,+}$ & 0.745 & 0.878 & FT & $\ket{-i,+}$ & 0.697 & 0.751 & nFT \\
$\ket{-,-}$ & 0.744 & 0.871 & FT & $\ket{-i,-}$ & 0.683 & 0.735 & nFT \\
\hline
& & & & $\ket{+i,+i}$ & 0.555 & 0.562 & nFT \\
& & & & $\ket{+i,-i}$ & 0.544 & 0.550 & nFT \\
& & & & $\ket{-i,+i}$ & 0.552 & 0.557 & nFT \\
& & & & $\ket{-i,-i}$ & 0.561 & 0.564 & nFT \\
\hline
\hline
\end{tabular}
\caption{Fidelities of 36 encoded Pauli states in the FBS code, measured with and without error detection using LQST.}
\label{table:fidelity}
\end{table}

\subsection{Logical state preservation}

\begin{figure}[h]
    \centering
    \includegraphics[width=0.9\textwidth]{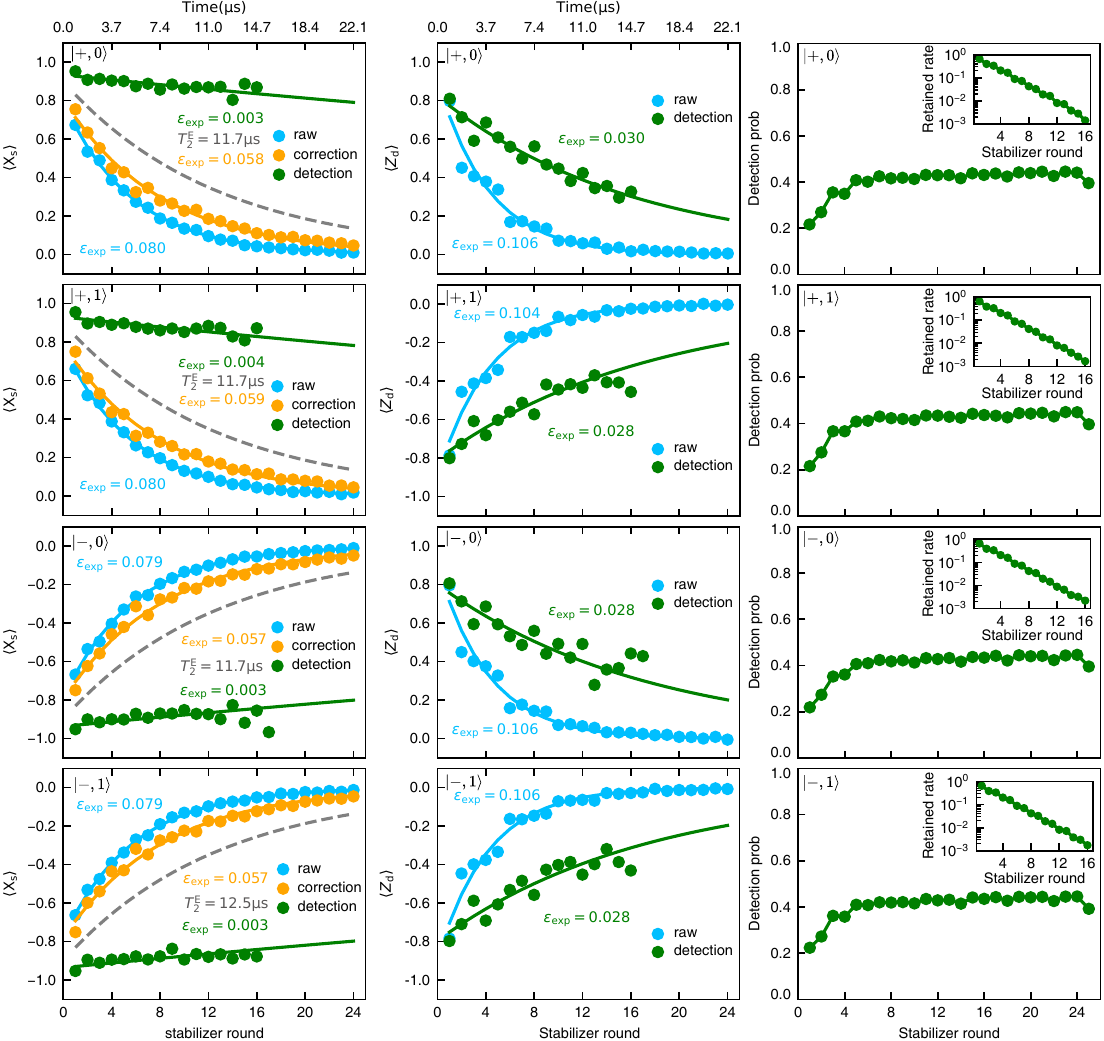}
    \caption{Logical state preservation in the FBS code.
    Multi-round state preservation results for the static qubit (left column) and the dynamical qubit (middle column), with different encoded states in each row. 
    Each stabilizer round lasts 920~ns, comprising 720~ns for readout pulses and resonator photon depletion, and 200~ns for single-qubit and two-qubit gates.
    The cyan, green, and orange dots correspond to data without post-processing, with error detection, and with error correction, respectively. 
    The average spin-echo dephasing time $T_2^{\rm E}=12.5$~$\mu$s of the physical qubits is shown as a dashed line for comparison.
    The corresponding error detection probabilities are shown in the right column, with the inset displaying the exponential decrease in the retained data rate during error detection, based on a total of 100,000 experimental shots.
    } \label{fig:exp04b_FBS_memory}
\end{figure}

The state preservation results for the static and dynamical qubit with encoded states $|+,0\rangle$, $|+,1\rangle$, $|-,0\rangle$, $|-,1\rangle$ are shown in Fig.~\ref{fig:exp04b_FBS_memory}, showing the measured expectation values for the $X_{\rm s}$ operator of the static qubit (left column) and the $Z_{\rm d}$ operator of the dynamical qubit (middle column). 

In the period-4 stabilizer measurements, each measurement result is compared to the result of the same stabilizer type from four rounds prior, except for the first four rounds and the last round. 
The measurement results of the first four stabilizer rounds are compared to the initial values determined by the encoding process. In the final round, equivalent stabilizers are calculated from the measurement outcomes of the data qubits and compared with the stabilizers of the same type from previous rounds.
The corresponding error detection probabilities are shown on the right of Fig.~\ref{fig:exp04b_FBS_memory}, with the insets displaying the exponential decrease in the retained data rate during error detection. 
Due to the lack of reset operations between stabilizer rounds, we perform XOR operations on the measurement results of the ancilla qubits during post-processing.

Taking the encoded state $|+,0\rangle$ as an example, the logical error rate of the static qubit is significantly reduced through error detection. Specifically, the error rate is suppressed from 8.0\% to 0.3\% by fitting the exponential decay of the expectation values. This improvement corresponds to an increase in the logical lifetime from 5.8~$\mu$s to 153~$\mu$s. The logical lifetime with error detection surpasses the average coherence time $T_2^{\rm E} = 12.5$~$\mu$s of the physical qubits. 
In contrast, the dynamical qubit exhibits a shorter logical lifetime due to reduced protection. With error detection, the logical error rate for the dynamical qubit is suppressed from 10.6\% to 3.0\%, corresponding to an increase in the logical lifetime from 4.3~$\mu$s to 15.3~$\mu$s. This difference arises because the static qubit requires three single-qubit errors to cause a logical error, whereas the dynamical qubit can experience a logical error with only two single-qubit errors.

\begin{figure}[!h]
    \centering
    \includegraphics[width=0.6\textwidth]{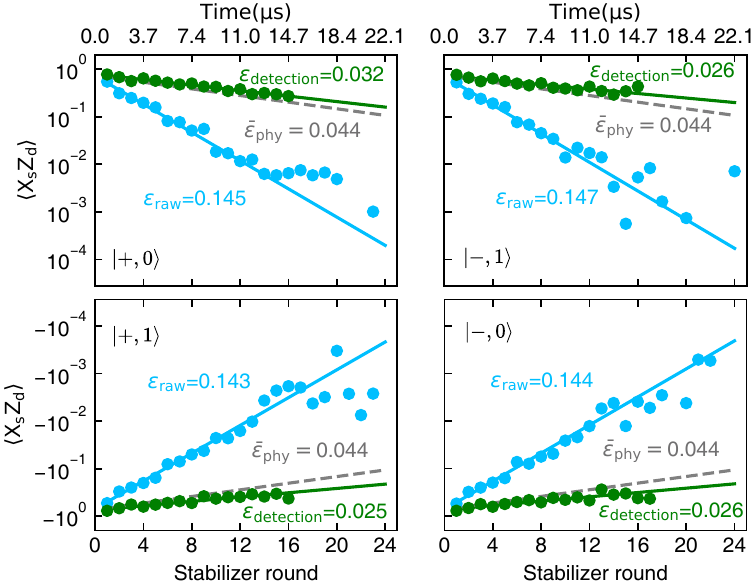}
    \caption{ 
    Logical state preservation for two-qubit states in the FBS code.
    Data are plotted on a logarithmic scale, with cyan (green) points representing raw (error-detected) results. 
    These data correspond to those in Fig.~\ref{fig:exp04b_FBS_memory} but are displayed here using a logarithmic y-axis to better illustrate the exponential decay behavior.
    } \label{fig:exp04b_FBS_memory_combine_prob}
\end{figure}

Beyond error detection, error correction can be performed on the static qubit alone, as it can be viewed as a BS code of distance 3. 
We decode the measured error syndromes using minimum-weight perfect matching algorithm~\cite{fowler2012surface}.
For the BS code, the MWPM decoder is applied independently to the two repetition codes: one derived from the $X$-type stabilizers to correct $Z$ errors, and the other from the $Z$-type stabilizers to correct $X$ errors. In our simulations, we implement a circuit-level noise model based on the error parameters listed in Table S4 and perform the MWPM decoding using the PyMatching package~\cite{gidney2021stim}.
For encoded state $|+,0\rangle$, error rate of the error-corrected static qubit is 5.8\%, corresponding to a logical lifetime of 8.3~$\mu$s.
However, the effectiveness of such error correction diminishes for the joint two-qubit states due to insufficient code distance and the correlation between errors in the two logical qubits.
We present error detection results for the two-qubit logical states in Fig.~\ref{fig:exp04b_FBS_memory_combine_prob}, using the same data as in Fig.~\ref{fig:exp04b_FBS_memory}. 

Notably, in Fig.~\ref{fig:exp04b_FBS_memory_combine_prob}, the fidelities at the initial points differ between the raw and error-detected data. This discrepancy arises because these initial points represent the logical fidelity of $X_{\rm s}Z_{\rm d}$ after one stabilizer round, not merely the state preparation fidelity. The complete circuit for these initial points comprises three parts: (1) encoding; (2) one round of stabilizer measurement (round A, shown in Fig.~1b of the main text); and (3) logical measurement in the $X_{\rm s}Z_{\rm d}$ basis (shown in Fig.~2(c) of the main text). As illustrated in Fig.~1(b) and Fig.~2(c) of the main text, stabilizers $S_X^{\text{A}}$ and $S_X^{\text{C}}$ are measured in round A and final logical measurement, respectively. These stabilizers enable the detection of a subset of physical errors occurring throughout the circuit. By post-selecting data in which no errors are detected, we obtain the error-detected data, which exhibits higher fidelity than the raw data. 

\subsection{Logical Bell states and Error Budget}

\begin{figure}[!h]
    \centering
    \includegraphics[width=0.95\textwidth]{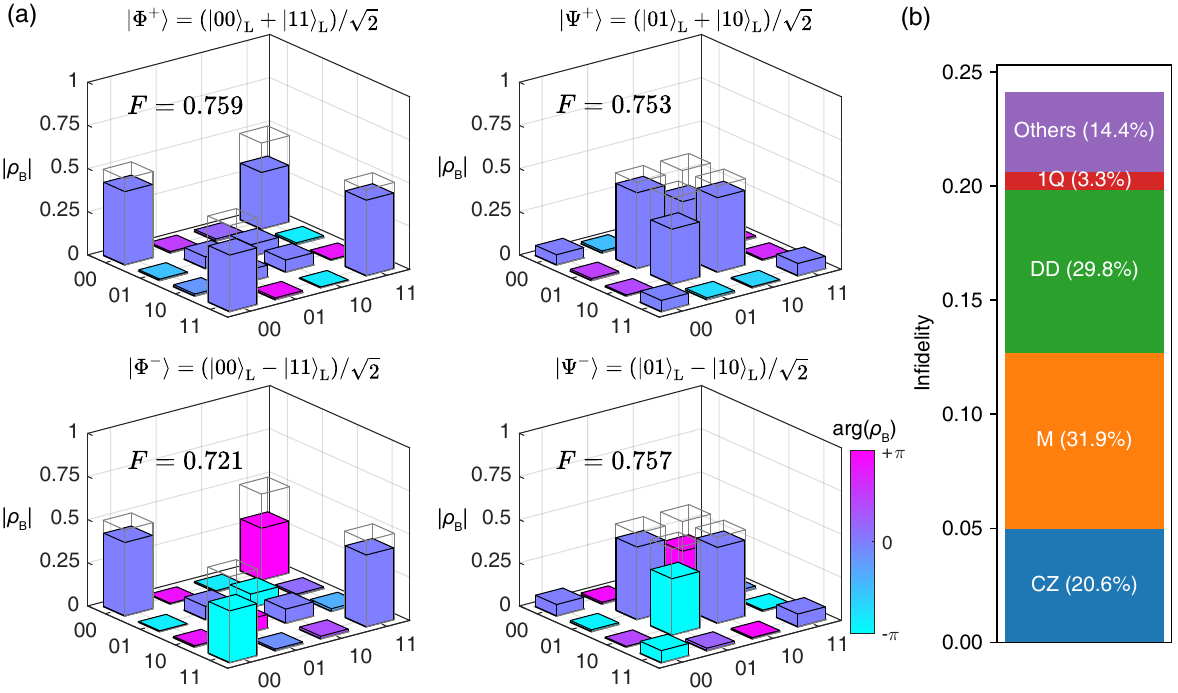}
    \caption{
    Logical Bell state characterion and error budget analysis.
    (a) LQST results of four Bell states $\{|\Phi^{+}\rangle, |\Psi^{+}\rangle, |\Phi^{-}\rangle, |\Psi^{-}\rangle\}$ generated from encoded states $\{|+,0\rangle, |+,1\rangle, |-,0\rangle, |-,1\rangle \}$ in the FBS code. The fidelities with error detection are shown. (b) Error budget decomposition for the $|\Phi^{+}\rangle$ state, quantifying contributions from single-qubit gates (1Q), CZ gates, measurement (M), dynamical decoupling (DD), and other unmodeled error sources.
    } \label{fig:exp05b_cnot_bell}
\end{figure}

We present the LQST results for four logical Bell states in Fig.~\ref{fig:exp05b_cnot_bell}(a). The measured fidelities of the states $\frac{1}{\sqrt{2}}(|0,0\rangle + |1,1\rangle)$, $\frac{1}{\sqrt{2}}(|0,1\rangle + |1,0\rangle)$, $\frac{1}{\sqrt{2}}(|0,0\rangle - |1,1\rangle)$, and $\frac{1}{\sqrt{2}}(|0,1\rangle - |1,0\rangle)$ with error detection are 0.759, 0.753, 0.721, 0.757, respectively.
Without error detection, the corresponding fidelities are 0.388, 0.369, 0.375, and 0.385.



To identify the main limitations and guide future improvements, we perform an error budget analysis for the logical Bell state $|\Phi^{+}\rangle$, following a methodology similar to Ref.~\cite{google_quantum_ai_exponential_2021}. First, we simulate the Bell-state generation circuit using Stim~\cite{gidney2021stim}, incorporating physical error rates for single-qubit gates (1Q), CZ gates, measurement (M), and dynamical decoupling (DD). The simulated fidelity of 0.794 agrees well with the experimental value of 0.759. We then compute the relative contributions of 1Q, CZ, DD, and M errors by evaluating the infidelity gradient with respect to each component at half the physical error rates. Each error source's infidelity contribution is given by the product of its physical error rate and the corresponding gradient weight. The results are summarized in Fig.~\ref{fig:exp05b_cnot_bell}(b) and Table~\ref{table_errorbudget}.

The error budget reveals that DD and measurement errors dominate the total infidelity, indicating that reducing errors during measurement—on both data and ancilla qubits—would most effectively enhance the logical fidelity. Additionally, approximately 14.4\% of the infidelity originates from sources beyond the 1Q/CZ/M/DD model, which we attribute to correlated errors induced by state leakage~\cite{mcewen2021removing}.

\begin{table}
\centering
\begin{tabular}{c | c | c | c | c }
\hline 
\hline 
~Component~ & ~Error rate~ & ~Weight~ & ~Infidelity contribution~ & ~Error percentage~ \\ 
\hline 
1Q     &  0.07\%  & 11.2 & 0.78\% & 3.30\%  \\ 
\hline 
CZ     &  0.97\%  & 5.13 & 4.97\% & 20.6\%  \\ 
\hline 
M      &  1.58\%  & 4.87 & 7.70\% & 31.9\%  \\ 
\hline 
DD     &  1.43\%  & 5.02 & 7.17\% & 29.8\%  \\ 
\hline 
others &          &      & 3.47\% & 14.4\%  \\ 
\hline
\end{tabular}
\caption{
Error budget analysis for the logical Bell state $\ket{\Phi^+}$, quantifying the contributions from different noise sources to the total infidelity. Components include single-qubit gates (1Q), CZ gates, measurement (M), dynamical decoupling (DD), and other unmodeled effects.
}
\label{table_errorbudget}
\end{table}

\clearpage

\bibliography{supp}  
\bibliographystyle{apsrev4-2}